\documentclass[acmtog]{acmart}
\AtBeginDocument{%
 \providecommand\BibTeX{{%
 \normalfont B\kern-0.5em{\scshape i\kern-0.25em b}\kern-0.8em\TeX}}}
\usepackage{multirow}
\usepackage{graphicx}
\usepackage{subcaption}
\usepackage{wrapfig}
\usepackage{hyperref}
\usepackage{url}

\setcopyright{acmcopyright}\acmJournal{TOG}
\acmYear{2022}\acmVolume{41}\acmNumber{4}\acmArticle{121}\acmMonth{7}
\acmDOI{10.1145/3528223.3530184}

\acmSubmissionID{786}

\begin{document}

\title{NeuralSound: Learning-based Modal Sound Synthesis with Acoustic Transfer}

\author{Xutong Jin}
\email{jinxutong@pku.edu.cn}
\affiliation{%
  \institution{School of Computer Science, Peking University}
 \country{China}
}

\author{Sheng Li}
\authornote{corresponding author\\Project URL: \href{https://hellojxt.github.io/NeuralSound}{hellojxt.github.io/NeuralSound}}
\email{lisheng@pku.edu.cn}
\affiliation{%
  \institution{School of Computer Science, Peking University}
  \country{China}
}

\author{Guoping Wang}
 \email{wgp@pku.edu.cn}
\affiliation{%
  \institution{School of Computer Science, Peking University}
  \country{China}
}

\author{Dinesh Manocha}
\email{dmanocha@umd.edu}
\affiliation{%
  \institution{University of Maryland at College Park}
 \country{U.S.A}
}

\ccsdesc[500]{Computing methodologies~Modeling and simulation}
\ccsdesc[300]{Applied
computing~Sound and music computing}

\begin{abstract}
We present a novel learning-based modal sound synthesis approach that includes a mixed vibration solver for modal analysis and a radiation network for acoustic transfer. Our mixed vibration solver consists of a 3D sparse convolution network and a Locally Optimal Block Preconditioned Conjugate Gradient (LOBPCG) module for iterative optimization. Moreover, we highlight the correlation between a standard numerical vibration solver and our network architecture. Our radiation network predicts the Far-Field Acoustic Transfer maps (FFAT Maps) from the surface vibration of the object. 
The overall running time of our learning-based approach for most new objects is less than one second on a RTX 3080 Ti GPU while maintaining a high sound quality close to the ground truth solved by standard numerical methods. We also evaluate the numerical and perceptual accuracy of our approach on different objects with various shapes and materials. 

\end{abstract}

\keywords{sound synthesis, deep learning, modal analysis, vibration, convolution networks, acoustic transfer, sound radiation}

\begin{teaserfigure}
\centering
 \includegraphics[trim={0cm, 0cm, 0cm, 0cm},clip,width=0.95\linewidth]{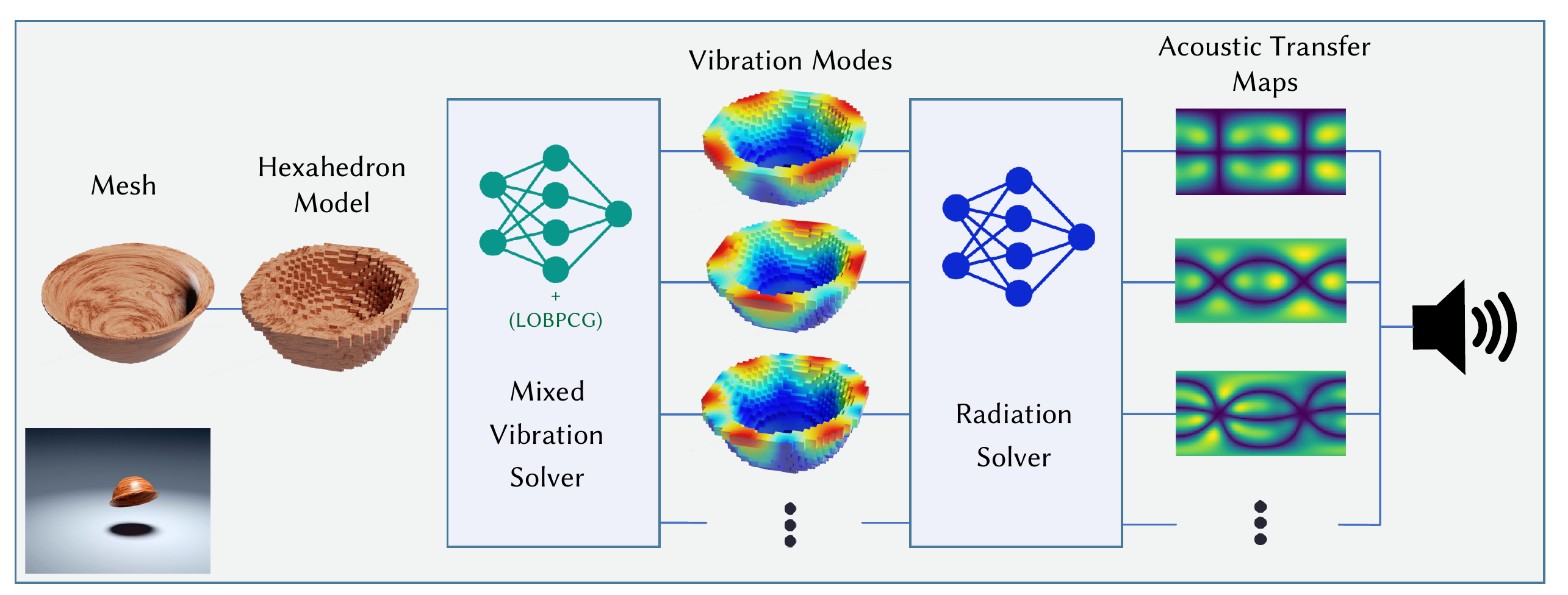}
\caption{
Learning-based approach for modal sound synthesis: We use neural networks to accelerate both modal analysis and acoustic transfer precomputation, and evaluate the performance on many new and unseen objects. Our approach can solve both vibration and radiation for plausible sound effects within one second per object on a GeForce RTX 3080 Ti GPU.
}
\label{fig:overview}
\end{teaserfigure}
\maketitle

\section{Introduction}
3D modal sound models are used for synthesizing physically-based rigid-body sounds in animation and virtual environments~\cite{van2001,modalsynthesis2002,Raghuvanshi2006, Bonneel2008,PAT,liu2020sound}. The standard pipeline generally requires expensive precomputation for sound vibration and radiation for each object in the scene. The sound vibration data is obtained by modal analysis~\cite{van1996, van2001}, which consists of an eigenmode matrix and eigenfrequencies that correspond to the surface motions and frequencies of some vibration modes. The sound radiation data corresponds to the acoustic transfer function~\cite{PAT}, which describes the radiation efficiency of each vibration mode at different spatial positions.


The standard numerical methods used for modal sound synthesis can be expensive, especially in terms of computing the sound radiation data~\cite{Interactive-AT, PAT}. For a given object, a numerical vibration solver~\cite{lanczos1950iteration} usually takes a few minutes~\cite{DeepModal, PAT, fracture}, and a numerical radiation solver such as Boundary Element Method (BEM)~\cite{bem_survey} can take several minutes to a few hours~\cite{PAT, kleinpat, fracture}. This makes it hard to use these methods in interactive scenarios with dynamically generated objects corresponding to shape modification or fractured objects.




\noindent{\bf Main Results:} We propose NeuralSound, a novel learning-based modal sound synthesis approach. 
It consists of a \emph{mixed vibration solver} and a \emph{radiation solver}. Our mixed vibration solver is a 3D linear sparse convolution network used to predict the eigenvectors of an object and an iterative optimizer that uses Locally Optimal Block Preconditioned Conjugate Gradient (LOBPCG) \cite{lobpcg,efficient_LOBPCG}. We highlight the connection between a 3D sparse convolution network and the matrix calculations of standard numerical solvers for modal analysis to motivate the design of our network.
We also present a neural network architecture based on the properties of BEM for acoustic transfer. Thus, our end-to-end radiation network encodes the surface displacement and frequency of a vibration mode and decodes it into the corresponding scalar-valued Far-Field Acoustic Transfer maps (FFAT Maps)~\cite{shell-ffat, kleinpat}, which compress the acoustic transfer function at the level-of-accuracy suitable for sound rendering. The total computation for sound vibration and radiation of an object can be completed within one second on a commodity GPU (RTX 3080Ti). 
Moreover, these two learning-based solvers can either be combined or used independently. In other words, a numerical vibration solver or a numerical radiation solver can be combined with our learning-based solver to design a sound synthesis system.

Our dataset is based on the ABC Dataset~\cite{abc_dataset} and uses $\sim$100K 3D models. We compare our mixed vibration solver with the standard numerical solution using different metrics. The result of a LOBPCG solver with a tight convergence tolerance is treated as the ground truth. Our mixed vibration solver obtains high accuracy (convergence error $\varepsilon$ < 0.01) using only 0.03s and always outperforms other standard numerical solvers within the same time budget. 
We compare our radiation solver with the standard numerical solution (BEM). Our approach achieves high accuracy (MSE of normalized FFAT Map  $\sim$ 0.06, and MSE of Log Norm $\sim$ 0.07) using only 0.04s. We observe $\sim$2000$\times$ speedup using our radiation solver over BEM in terms of synthesizing plausible acoustic transfer effect.

We conduct a user study to evaluate our sound quality compared to two baselines in five different scenes. The statistical results (mean value and standard deviation) indicate that our approach, including both the mixed vibration solver and the radiation solver, shows very high fidelity to the ground truth.


\section{Related works}

\paragraph{Modal Sound Synthesis} Modal sound synthesis is a technique that has been used to synthesize sounds of rigid bodies~\cite{cook1995,van2001,modalsynthesis2002,Raghuvanshi2006}. These methods compute the vibration modes of a 3D object using eigendecomposition as a preprocessing. Many methods use 3D objects' pre-computed eigendata to render runtime sound, e.g., reducing the computational complexity by approximations~\cite{Bonneel2008}. Other methods use complex modal sound synthesis models to simulate sounds such as knocking, sliding, and friction models~ \cite{van2001}, acceleration noise synthesis models~\cite{Acceleration-Noise}, accurate damping models~\cite{sterling2019}, contact models \cite{contact} or data-driven approximations~\cite{ren2013example, pai2001scanning}.

\paragraph{Acoustic Transfer} 
Directly adding the vibration modes does not result in high-fidelity sound effects, as that process lacks the modal amplitude variations and spatialization effects due to acoustic wave radiation~\cite{kleinpat}. In addition to BEM~\cite{bem_survey}, precomputed acoustic transfer~\cite{PAT} methods are used to more accurately model the sound pressure at the listener's position.
Other techniques use a single-point multipole expansion with higher-order sources~\cite{contact,fracture,eigenmode-compression,rungta2016syncopation}, inter-object transfer functions~\cite{mehra2013wave}, or Far-Field Acoustic Transfer (FFAT) Maps~\cite{shell-ffat, kleinpat}. Since the FFAT map is a simple,  efficient representation of the transfer function, we also use scalar-valued FFAT Maps defined in KleinPAT~\cite{kleinpat} in our approach. 

\paragraph{Precomputation Speedup} Some methods focus on reducing the time-consuming precomputation of modal sound synthesis and acoustic transfer functions.  
Li et al.~\shortcite{Interactive-AT} proposed a method to enable interactive acoustic transfer evaluation without recalculating the acoustic transfer when changing an object's material. 
KleinPAT~\cite{kleinpat} can accelerate the precomputation of acoustic transfer on the GPU by packing-depacking modes and using a time-domain method. 
A deep neural network~\cite{DeepModal} was proposed to synthesize the sound of an arbitrary rigid body  being hit at runtime and it predicts the amplitudes and frequencies of sound directly. However, this approach cannot model acoustic transfer. Moreover, it may incur a severe loss of accuracy because it treats all the modes within a frequency interval as one mode.

\paragraph{Learning-based Approaches} 
The computation of modal vibration and radiation is highly dependent on the geometric shape of the object, where solving for shape-related features is the key issue. Many methods have been proposed to utilize three-dimensional geometric features for learning, including perspective-based methods, which regard the combination of 2D image features from different viewing perspectives as the 3D geometry feature~\cite{perspective1}, voxel-based methods~\cite{voxel1}, which regard an object as a 3D image by hexahedral meshing, and point cloud-based methods~\cite{point1, DBLP:journals/corr/abs-2105-08177}, which represent the model as a point cloud. 
 
\section{Background }
\label{sec:pre}
We first provide the background of the 3D physically-based modal sound synthesis as well as the acoustic transfer function for high-quality sound effect.

\subsection{Vibration Solver}
\label{background_vibration}

We begin with the linear deformation equation for a 3D linear elastic dynamics model~\cite{shabana1991theory} that is commonly used in rigid-body sound synthesis~\cite{fracture, modalsynthesis2002, PAT}:
\begin{equation}\mathbf{M} \ddot{\mathbf{u}}+\mathbf{C} \dot{\mathbf{u}}+\mathbf{K}\mathbf{u} = \mathbf{F}(t)\ ,
\label{eq:ModalAnalysis}
\end{equation}
where $\mathbf{u} $ is nodal displacements, and
$\mathbf{M},\ \mathbf{C}=\alpha \mathbf{M}+\beta \mathbf{K},$  and $\mathbf{K} $
represent the mass, Rayleigh damping, and stiffness matrices,
respectively. $\alpha, \beta$ are Rayleigh damping coefficients. $\mathbf{F}(t)$ represents the external nodal forces. The generalized eigenvalue decomposition as
\begin{equation}
    \mathbf{K U}= \mathbf{M} \mathbf{U} \mathbf{\Lambda}\ 
    \label{eq:generalized_eigen_decomposition}
\end{equation}
is required first to solve the linear system, where $\mathbf{U}$ is the eigenmode matrix (consists of eigenvectors) and $\Lambda$ is the diagonal matrix of eigenvalues. Then the system can be decoupled as:
\begin{equation}\ddot{\mathbf{q}}+(\alpha+\beta \mathbf{\Lambda}) \dot{\mathbf{q}}+\mathbf{\Lambda} \mathbf{q}=\mathbf{U}^{T} \mathbf{F}(t)\ ,
\end{equation}
where $\mathbf{q} $ satisfies $\mathbf{u}=\mathbf{U}\mathbf{q} $. The solution to the equation is a bank of damped sinusoidal waves corresponding to each mode. The generalized eigenvalue decomposition $\mathbf{K U}= \mathbf{M} \mathbf{U} \mathbf{\Lambda}$ is the core of modal analysis. Therefore, a generalized eigenvalue decomposition solver is required (e.g., Lanczos method~\cite{lanczos1950iteration}, LOBPCG~\cite{lobpcg,efficient_LOBPCG}).


\subsection{Acoustic Transfer Solve}
\label{background_acoustic_transfer}

Acoustic transfer function $p_{i}(\mathbf{x})$ describes the sound pressure in space (at position $\mathbf{x}$) generated by the ith mode with unit amplitude. 
A radiation solver is required to solve the acoustic transfer function from the surface displacement and frequency of a mode, where the surface displacement is computed from the eigenmode matrix $\mathbf{U}$ solved in Sec. \ref{background_vibration}, and the frequencies are computed from the eigenvalues $\Lambda$. BEM is a standard method used to solve this acoustic transfer problem. After solving the acoustic transfer function, compression methods are needed to represent the function for runtime sound rendering. These methods include Equivalent-Source Representations~\cite{PAT,mehra2013wave}, Multipole Source Representation~\cite{fracture}, Far-Field Acoustic Transfer (FFAT) Maps \cite{shell-ffat}, etc. In this paper, we choose FFAT Maps as our compression method.

\subsection{Sound Synthesis}
\label{background_sound_synthesis}

To synthesize modal sound at runtime, the eigenvalue matrix $\mathbf{\Lambda} = diag\{ \lambda_1,  \lambda_2, ..., \lambda_k\}$, eigenmode matrix $\mathbf{U}$, and acoustic transfer function $\mathbf{P}(\boldsymbol{x}) = \{p_{i}(\boldsymbol{x}) | i = 1,...,k\}$ are precomputed and stored~\cite{eigenmode-compression}.
The external nodal force $\mathbf{F}(t)$ is projected into the modal subspace by $\mathbf{U}^T\mathbf{F}(t) = \{f_0(t), f_1(t), ..., f_k(t)\}^T$, then the sound waveform $w(\boldsymbol{x},t)$, i.e., the sound pressure at position $\boldsymbol{x}$ and time $t$, is solved as \cite{sigcourse}:
$$
w(\boldsymbol{x},t) = \sum_{i = 1}^{k} p_{i}(\boldsymbol{x}) \int_{0}^{t} \frac{f_i(\tau)}{\omega'_i} e^{-\xi _i \omega_i(t-\tau)} \sin \left( \omega'_i(t-\tau) \right) \mathrm{d} \tau,
$$
where $\omega_i = \sqrt{\lambda_i}$ is the undamped natural frequency of $i$th mode, $\xi_i =  \frac{\alpha + \beta \lambda_i}{2 \omega_i}$ is the dimensionless modal damping factor of $i$th mode, and $\omega'_i = \omega_i \sqrt{1-\xi_i^2}$ is the damped natural frequency of $i$th mode.

\subsection{Network Architectures for Modal Sound Synthesis}
Modal analysis and acoustic transfer precomputations can be expensive, making the current numerical solvers too slow for interactive applications. In contrast to these solvers, the inference process of a neural network can be completed in a very short time (even milliseconds) on current GPUs. Therefore, a learning-based approach to resolving modal sound synthesis can inherit the advantage of high efficiency.

Our approach exploits the intrinsic correspondence between the convolution neural network and physically-based sound synthesis methods. Specifically, local connections and shift-invariance are two characteristics of convolution neural networks (CNNs). We also observe similar characteristics for (i) the assembled matrix (i.e., stiffness matrix) in modal analysis and (ii) the interaction/convolution matrix in the BEM for acoustic transfer~\cite{bem_survey}. Our approach is also inspired by the fact that multi-scale structure in convolution neural networks (e.g., in ResNet~\cite{resnet} and U-Net~\cite{Unet}) has also been applied to (i) the multigrid method in linear solvers and (ii) the hierarchical strategy in the fast multipole method (FMM)~\cite{fmmbem}.

We design two neural networks that coincide with the fundamentals of numerical solvers for modal analysis and acoustic transfer but can predict approximate results much faster. This is the motivation behind our learning-based approach. Specifically, our approach includes (i) a mixed vibration solver for modal analysis and (ii) a radiation solver for acoustic transfer (see \autoref{fig:overview}). For our mixed vibration solver, we highlight its correspondence with the numerical solvers and present its details in Sec.~\ref{sec:mixed_vibration_solver}. The correspondence with BEM and the details of our radiation solver are given in Sec.~\ref{sec:acoustic}.

\section{Vibration Solver: Learning Eigenvector Approximation}
\label{sec:mixed_vibration_solver}
In this section, we propose a self-supervised learning-based solver to resolve vibration problems introduced in Sec.~\ref{background_vibration}, i.e., to resolve the core generalized eigenvalue decomposition (\autoref{eq:generalized_eigen_decomposition}). We explain the design of our vibration solver, present the architecture, and describe training our 3D sparse network by reducing residual-based error instead of using data generated from a numerical solver. 

\subsection{Network and Matrix Computations}
\label{sec:consistency}

The assembled matrix and its inverse are key components in numerical solvers for generalized eigenvalue decomposition in modal analysis. We first analyze the intrinsic connections between the assembled matrix and the 3D sparse convolution as well as the inverse assembled matrix and the 3D sparse U-Net. 

\subsubsection{Assembled Matrix and 3D Sparse Convolution}
We use 3D sparse convolutions~\cite{mink, sparse1, sparse2} as the basic components of our network. 3D sparse convolution can be applied to sparse tensors of any shape due to the shift-invariance of convolution kernel. In a $3\times3\times3$ sparse convolution, the convolution kernel defines the linear relationship between the output feature of each voxel and the input features of neighboring voxels (including itself). In an assembled matrix, the element matrix defines the linear relationship between the output displacement of each vertex and the input displacements of neighboring vertices (including itself) in a voxel. So, there is an intuitive connection between 3D sparse convolutions and assembled matrices. Assuming that the element matrix is fixed (voxel size and material are fixed), the assembled matrix-vector multiplication can be represented by a corresponding $3 \times 3 \times 3$ sparse convolution (see bottom-right of \autoref{fig:eigen_net}). We provide detailed analysis and experimental validation in Appendix \ref{sec:valid_network}.

\subsubsection{Inverse Assembled Matrix and Sparse U-Net}
Using a multigrid method~\cite{multigrid} can accelerate solving inverse matrix-vector multiplication through a hierarchy of discretizations, which refine the solution first from a fine to coarse model and then from coarse to fine. Each correction operation corresponds to matrix-vector multiplication and addition, equivalent to a 3D sparse convolution with bias if the involved matrix is an assembled matrix. Therefore, there is an intuitive connection between a 3D sparse linear U-Net~\cite{Unet} and the inverse of an assembled matrix. We provide experimental validation in Appendix \ref{sec:valid_network} to show that 3D sparse linear U-Net can be a good approximation of an assembled matrix inverse.

\subsubsection{Sparse Linear U-Net for Eigenvector Approximation}
\label{U_Net_Eigenvector_relationship}

The matrix  $\mathbf{K}^{-1}\mathbf{M}$ can be used to span the standard Krylov space where the Rayleigh–Ritz method~\cite{lobpcg, ritz1, ritz2} is applied to resolve approximate eigenvectors within this space. We explain the feasibility and rationality of such a process used for modal analysis with example in Appendix \ref{example_solve_eigenvectors}.
In principle, a linear neural network with certain parameters can represent a linear mapping. Based on the above analysis, a 3D sparse linear U-Net can represent a linear mapping that approximates an assembled matrix (or inverse matrix). Therefore, we conjecture that the U-Net can be used to approximate a polynomial of an assembled matrix (or inverse matrix) when the network with multiple layers is deep enough.
A 3D sparse linear U-Net with different parameters can represent various linear mappings. All these possible linear mappings make up the feasible domain of this U-Net. Like Krylov space, U-Net representation of the linear mappings can span a subspace and employ Rayleigh-Ritz method for eigenvector approximations. Based on the large feasible domain of deep neural networks, we can train the U-Net to search for the best possible linear mapping in its feasible domain rather than to simulate a specific one like $\mathbf{K}^{-1}\mathbf{M}$ of the krylov space. 


\begin{figure*}[htb]
\centering
 \includegraphics[width=0.9\linewidth]{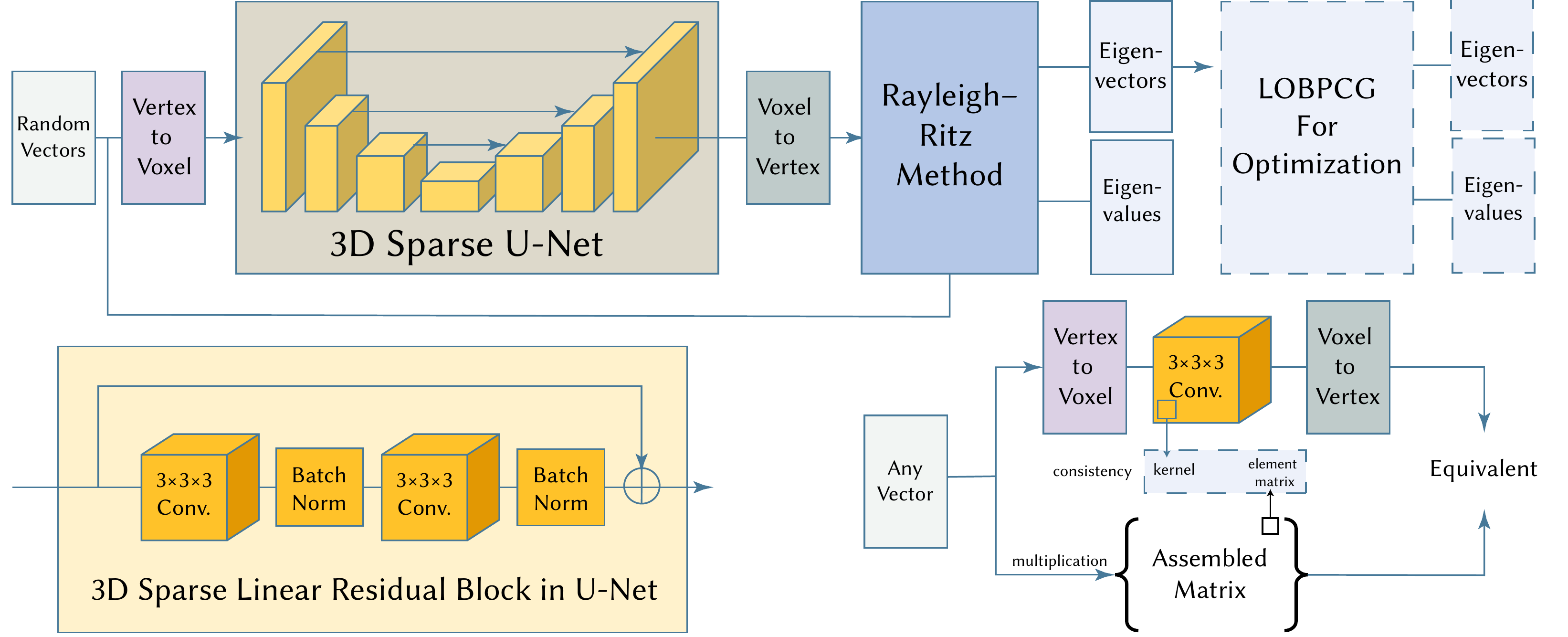}
\caption{
(\textbf{Top}) Architecture of our vibration solver. $k$ random initial vectors are fed into the 3D sparse U-Net. The output vectors are concatenated with random vectors and converted into the approximate eigenvectors and eigenvalues using the Rayleigh-Ritz method. LOBPCG will optimize The approximate eigenvectors to further reduce errors. (\textbf{Bottom left}) Linear residual block in our U-Net.  (\textbf{Bottom right}) The connection behind our network design: a $3\times3\times3$ 3D sparse convolution is equivalent to an assembled matrix with a corresponding element matrix.
}
\label{fig:eigen_net}
\end{figure*}

\subsection{Vibration Network Architecture}
\label{sec:eigen_network_structure}
As discussed above, we design a 3D sparse linear U-Net to transform some random initial vectors into output vectors. The output vectors span a subspace, and the Rayleigh–Ritz method is used to compute the approximate eigenvectors and eigenvalues in this subspace. We use a self-supervised training strategy without any labels or groundtruth data. The convergence error of the approximate eigenvectors solved by Rayleigh–Ritz method is used as the loss function.

The overall pipeline is shown in \autoref{fig:eigen_net}. Our U-Net consists of several linear residual blocks. The down-sampling layer in the U-Net is a 3D sparse convolution with a stride of 2. The upsampling layer in the U-Net is a 3D sparse transposed convolution with a stride of 2.

\subsubsection{Projection between Vertex \& Voxel}
\label{sec:projection}
First, we introduce the mechanism to ensure the accurate correspondence between 3D sparse tensors and eigenvectors. Voxels in the network correspond to hexahedrons in the finite element model. An eigenvector represents the displacement of vertices in the direction of $x,y,z$ caused by a unit vibration mode. A 3D sparse tensor in the neural network represents the feature of voxels. We design a projection from vertex representation to voxel representation and vice versa. The vertex to voxel projection concatenates all the features (displacement in $x, y, z$) of the current voxel's eight vertices. Therefore, the feature number of each voxel after projection is 24. For each vertex, 3 of the 24 features of adjacent voxels belong to this vertex. The voxel to vertex projection averages the corresponding three features of all adjacent voxels.

\subsubsection{Training Process}

We denote our neural network as $g$ and its optimizable parameters as $\mathbf{\theta}$. There are $N$ voxel models in the training dataset, each consisting of $k$ randomly initialized vectors $\mathbf{x}^{i} = \left(\mathbf{x}^i_1,\mathbf {x}^i_2,...,\mathbf{x}^i_k\right)$, which represent the randomly initialized nodal displacements. The goal is to find an optimal set of network parameters $\hat{\mathbf{\theta}}$ to minimize the average loss function:
\begin{equation}
    \hat{\mathbf{\theta}}=\underset{\mathbf{\theta}}{\operatorname{argmin}} \frac{1}{N} \sum_{i=1}^{N} \ell\left(\text{Rayleigh-Ritz}\left(g(\mathbf{x}^{i} ; \boldsymbol{\theta}),\mathbf{x}^{i}\right)\right) \ ,
\label{training}
\end{equation}
where $g(\mathbf{x}^{i} ; \boldsymbol{\theta}) = \left( g(\mathbf{x}^{i}_1 ; \boldsymbol{\theta}),g(\mathbf{x}^{i}_2 ; \boldsymbol{\theta}),...,g(\mathbf{x}^{i}_k ; \boldsymbol{\theta}) \right)$ represents $k$ output vectors from the network, and $\ell$ is the loss function. All computation are differentiable, so the common gradient optimization method can be employed. The loss function and the Rayleigh-Ritz method will be described in detail in the next subsection. 

\subsubsection{Rayleigh-Ritz Method \& Loss Function}

The Rayleigh-Ritz method is used to find the approximate eigenvectors and eigenvalues in a subspace:
\begin{equation}
    (\mathbf{S}^T\mathbf{K}\mathbf{S})\hat{\mathbf{U}} = (\mathbf{S}^T\mathbf{M}\mathbf{S})\hat{\mathbf{U}} \hat{\mathbf{\Lambda}}\ ,
\label{eq:rayleigh-ritz}
\end{equation}
where $\mathbf{S}$ is a set of basis of the linear subspace $\mathcal{S} = \operatorname{span}\{g(\mathbf{x}^{i} ; \boldsymbol{\theta}),\mathbf{x}^{i}\}$ and $\mathbf{K}, \mathbf{M}$ are stiffness matrix and mass matrix, respectively. The Ritz pairs $(\hat{\mathbf{\Lambda}}, \mathbf{S}\hat{\mathbf{U}})$ are approximations to the eigenvalues and eigenvectors in the original problems. It turns out that $k \ll 3N$, where $N$ is the number of vertices. Therefore, a solution to the generalized eigenvalue problem in subspace (\autoref{eq:rayleigh-ritz}) is much faster than directly solving the original generalized eigenvalue problem in modal analysis (\autoref{eq:generalized_eigen_decomposition}). 

The vectors in $g(\mathbf{x}^{i} ; \boldsymbol{\theta})$ may be linearly dependent. In this case, the dimension of the spanned subspace $\mathcal{S} = \operatorname{span}\{g(\mathbf{x}^{i} ; \boldsymbol{\theta})\}$ is less than $k$ and the number of eigenvectors that can be solved in the Rayleigh-Ritz process is also less than $k$. To ensure $k$ eigenvectors, we supplement random vectors $\mathbf{x}^{i}$ into subspace $\mathcal{S}$ as $\mathcal{S} = \operatorname{span}\{g(\mathbf{x}^{i} ; \boldsymbol{\theta}),\mathbf{x}^{i}\}$ because $\mathbf{x}^{i}$ are linearly independent. 

A numerical instability issue may occur in the Rayleigh-Ritz process in \autoref{eq:rayleigh-ritz}. This is due to the fact that the projection $\mathbf{S}^T\mathbf{M}\mathbf{S}$ can be ill-conditioned or rank deficient \cite{efficient_LOBPCG}. We use the SVQB algorithm \cite{efficient_LOBPCG} to convert the generalized eigenvalue problem of \autoref{eq:rayleigh-ritz} into a standard eigenvalue problem, which can be solved by numerically stable method. 

Since Rayleigh-Ritz method is critical to our approach, we provide the validation experiments in Appendix \ref{sec:valid_ritz}.

Loss function is also a critical part of the self-supervised training strategy. When the Ritz pairs $(\hat{\mathbf{\Lambda}}, \mathbf{S}\hat{\mathbf{U}})$ is solved, we denote the approximate eigenvalues as $\hat{\mathbf{\Lambda}} = diag\{\hat \lambda_1, \hat \lambda_2, ..., \hat \lambda_k\}$ and the approximate eigenvectors as $ \mathbf{S}\hat{\mathbf{U}} = [\hat{ \mathbf{v_1}},\hat{\mathbf{v_2}},...,\hat{\mathbf{v_k}}]$. The loss function is defined as: 
\begin{equation}
   \ell(\hat{\mathbf{\Lambda}}, \mathbf{S}\hat{\mathbf{U}}) = \frac{1}{k} \sum_{i=1}^k \frac {||\mathbf{K}\hat{\mathbf{v_i}} - \hat\lambda_i \mathbf{M} \hat{\mathbf{v_i}}||_2}  {(||\mathbf{K}||_2 + |\hat\lambda_i|||\mathbf{M}||_2)||\hat{\mathbf{v_i}}||_2} + \gamma \frac{1}{k} \sum_{i=1}^k |\hat\lambda_i| \ ,
    \label{eq:loss}
\end{equation}
where $\gamma$ is a manually set hyper-parameter. The left item of \autoref{eq:loss} is the residual-based error defined in the convergence criterion of LOBPCG~\cite{efficient_LOBPCG}. Reducing this residual-based error increases the accuracy of the solved eigenvectors. The right item of \autoref{eq:loss} is the average of the solved eigenvalues. Reducing the average eigenvalue can facilitate resolving the first $k$ smallest eigenvalues for modal analysis.

Our U-Net is trained with a fixed material and voxel size. The eigenvectors and eigenvalues of an object with different materials or voxel sizes can be solved by linearly scaling the results obtained by our vibration solver. For more details about the scaling, please refer to~\citet{fracture, DeepModal}.

\subsubsection{Warm starting to LOBPCG}

In principle, a neural network-based method does
not have the characteristic of convergence through iterations like traditional numerical solutions. Therefore, the predicted values from our vibration solver will inevitably produce some errors. To reduce the accuracy loss and further improve sound quality, we design a mixed vibration solver that uses the results of our learning-based module to warm-start an LOBPCG solver~\cite{lobpcg, efficient_LOBPCG}.

These two modules can be integrated naturally because the output of our network is consistent with the representation of standard finite element model (see Sec.~\ref{sec:consistency} and \ref{sec:projection}). 

Since our network can yield outputs close to the actual eigenvectors, the warm-started LOBPCG solver tends to converge quickly. In other words, our mixed vibration solver can obtain more accurate results quickly, as shown in Sec.~\ref{sec:results}.

\section{Radiation Solver: learning FFAT Map}
\label{sec:acoustic}

We propose a learning-based radiation solver for the acoustic transfer problem introduced in Sec. \ref{background_acoustic_transfer}. Our radiation network predicts the acoustic transfer function from the surface displacement and frequency of a vibration mode. We select the scalar-valued FFAT Map as the compressed representation of the acoustic transfer function. The scalar-valued FFAT Map compresses the  acoustic transfer function of a mode into a 2D image as \cite{shell-ffat} :
\begin{equation}
\frac{\psi_{i}(\theta, \phi)}{r} \approx |p_i(\mathbf{x})| \ ,
\label{eq:ffat_map}
\end{equation}
where $\theta, \phi$ are spherical coordinates with the center of the object as the origin, and $r$ is the distance from $\mathbf{x}$ to the origin. The pixel value of $(\theta, \phi)$ position in the 2D image is $\psi_{i}(\theta, \phi)$. We provide the details of the scalar-valued FFAT Map in Appendix \ref{sec:formulation_acoustic}. 
In the following subsections, we first highlight the connection between the convolution neural network and a standard numerical method, i.e., BEM. Next, we introduce our end-to-end radiation network for fast precomputation of the acoustic transfer function.

\begin{figure*}[ht]
\centering
\includegraphics[width=0.9\linewidth]{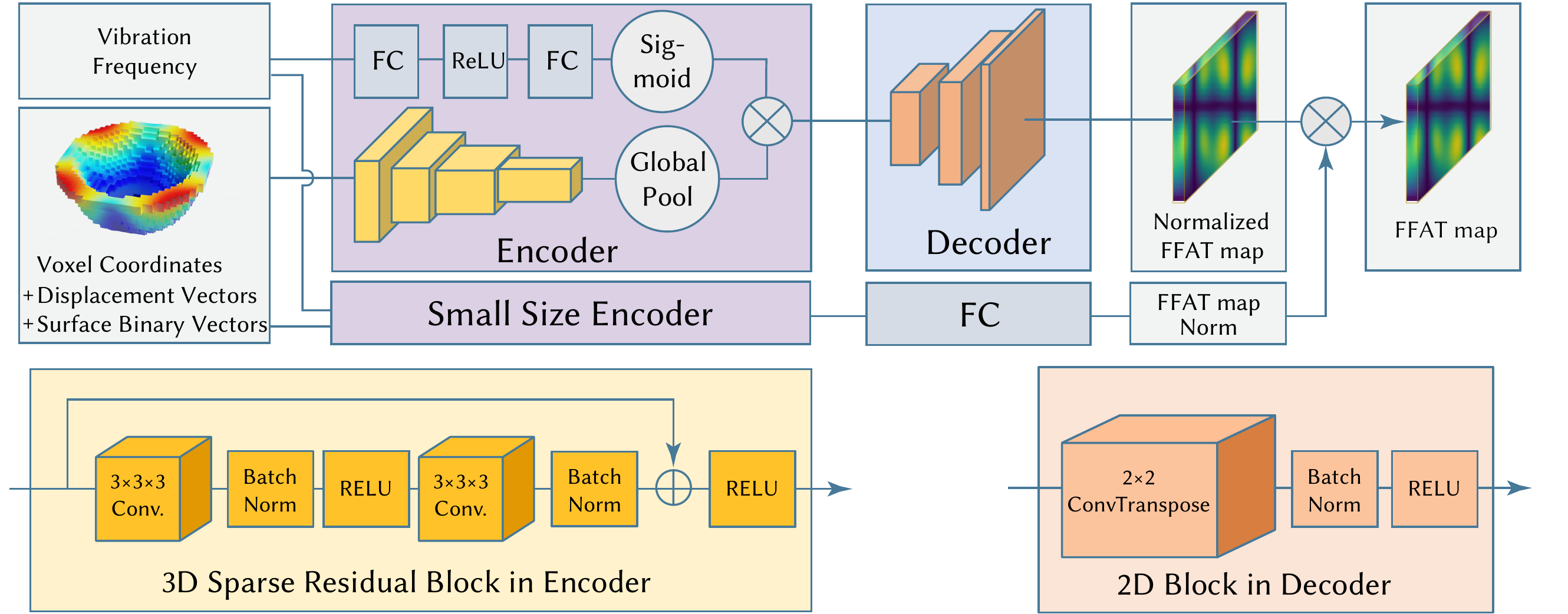}
\caption{
(\textbf{Top}) Architecture of our radiation solver. It consists of two branches for normalized FFAT Map and FFAT Map norm. In each branch, the encoder consists of a 3D sparse ResNet for encoding the features of surface voxels and two fully connected layers for encoding vibration frequency. Two encoder parts fuse by multiplying their results. The ResNet consists of several residual blocks (\textbf{Bottom left}). The decoder for normalized FFAT Map consists of several 2 $\times$ 2 transpose convolutional blocks (\textbf{Bottom right}). In addition, the decoder for FFAT Map norm is a fully connected layer. }
\label{fig:transfer_net}
\end{figure*}

\begin{figure}[ht]
\centering
\includegraphics[trim={1.7cm 0cm 3.5cm 1.3cm},clip,width=1.0\linewidth]{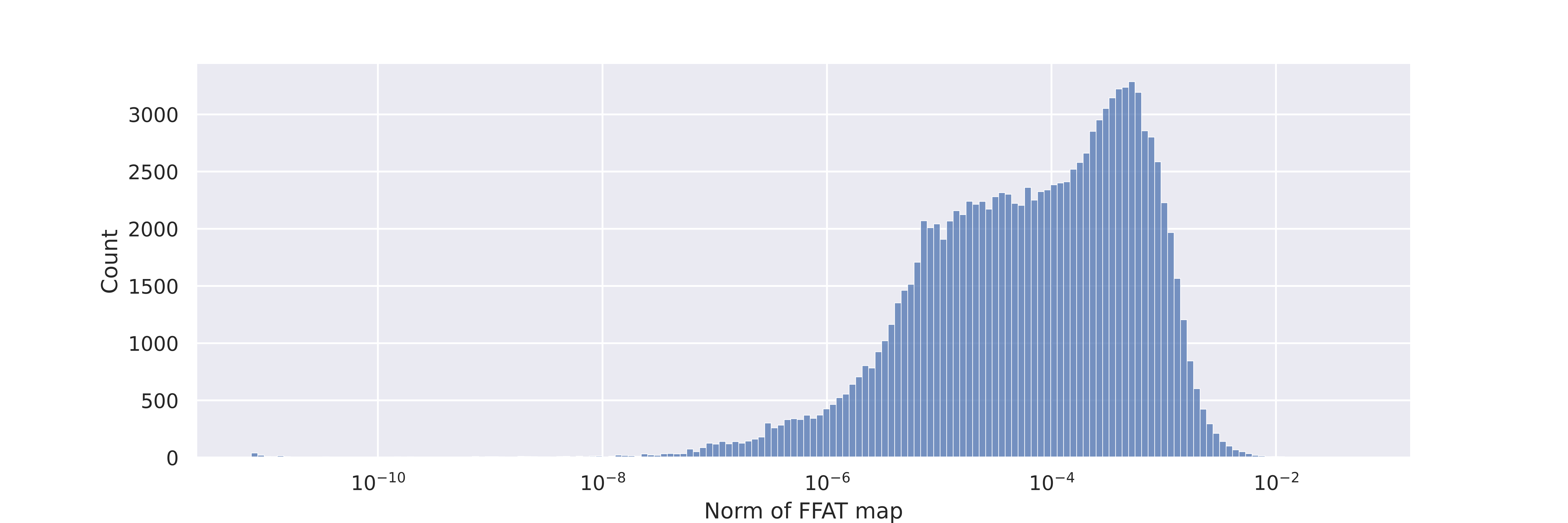}
\caption{Distribution of FFAT Map norms in our dataset. The surface vibration vectors are normalized to a unit norm before being fed to BEM to solve FFAT Maps. 
We trained our network to predict the norm in log scale due to its distribution's smoothness. }
\label{fig:transfer_norm}
\end{figure}

\subsection{Network and Radiation Solver}
\label{sec:acoustic_connection}
BEM is a standard method used to compute the acoustic transfer function. We first analyze the intrinsic connections between the neural network and the BEM, and use that to design our radiation network.

Assume that the surface acoustic transfer value $\mathbf{p}_{surf}$ and its normal derivative $\partial_{\boldsymbol{n}} \mathbf{p}_{surf}$ on a surface are known. The acoustic transfer values $\mathbf{p}_{outer}$ on an outer sphere is needed for FFAT Map (see \autoref{fig:ffat} in Appendix) and can be computed by the potential operators~\cite{fmmbem, betcke2021bempp}:
\begin{equation}
    \mathbf{p}_{outer} = \mathcal{V}\partial_{\boldsymbol{n}} \mathbf{p}_{surf} - \mathcal{K}\mathbf{p}_{surf} \ ,
\end{equation}
where $\mathcal{V}, \mathcal{K}$ are the single layer potential operator and the double layer potential operator, respectively. Fast Multipole Method (FMM)~\cite{fmmbem} can be applied to compute the projection ($\mathcal{V}$ and $\mathcal{K}$) from the object surface to the outer sphere. The element-to-element interactions in the conventional BEM can be analogized to cell-to-cell interactions within a hierarchical tree structure of cells containing groups of elements. Therefore, there is an intuitive connection between a neural network with a downsampling-upsampling architecture and this sound radiation process.

Specifically, the downsampling part of our network is inspired by the particles to multipole (P2M) and multipole to multipole (M2M) processes in FMM. We observe two important features of FMM: (i) shift-invariance, since interactions depend on the relative position between elements; (ii) spatial locality, since nearby elements are aggregated into cells. They are also features of 3D sparse convolution network, as described in~\cite{mink}. Therefore, we apply 3D sparse convolution as the downsampling part of our network. 

Similarly, the upsampling part is inspired by the local to local (L2L) and local to particles (L2P) computations in FMM. Shift-invariance and spatial locality are also applicable. Therefore, we apply 2D transpose convolution as the upsampling part of our network. 

It turns out that  $\partial_{\boldsymbol{n}} \mathbf{p}_{surf}$ can be computed from the surface vibration, whereas $\mathbf{p}_{surf}$ is still unknown. Without considering the fictitious frequency~\cite{fmmbem,Interactive-AT},  $\mathbf{p}_{surf}$ can be solved from the linear equations of  the conventional boundary integral equation (CBIE)~\cite{fmmbem, betcke2021bempp}:
\begin{equation}
    (\frac 1 2 \mathbf{I} + \mathbf{K}) \mathbf{p}_{surf}= \mathbf{V} \partial_{\boldsymbol{n}} \mathbf{p}_{surf} \ ,
\end{equation}
where $\mathbf{I}, \mathbf{V}, \mathbf{K}$ are identity operator, single layer boundary operator, and double layer boundary operator, respectively. $\mathbf{V}$ is derived from Green’s function $G(\boldsymbol{x} ; \boldsymbol{y})=\frac{e^{i k\|\boldsymbol{x}-\boldsymbol{y}\|}}{4 \pi \|\boldsymbol{x}-\boldsymbol{y}\|}$, and $\mathbf{K}$ is derived from $\frac{\partial G(\boldsymbol{x}; \boldsymbol{y})}{\partial \boldsymbol{n}(\boldsymbol{y})}$. $G(\boldsymbol{x} ; \boldsymbol{y})$ and $\frac{\partial G(\boldsymbol{x}; \boldsymbol{y})}{\partial \boldsymbol{n}(\boldsymbol{y})}$ are inversely proportional to the relative distance from $\boldsymbol{x}$ to $\boldsymbol{y}$, which shows the spatial locality and shift-invariance between the elements of $\partial_{\boldsymbol{n}} \mathbf{p}_{surf}$ and  the elements of $\mathbf{p}_{surf}$.
We infer that 3D sparse convolution can also handle the translation from $\partial_{\boldsymbol{n}} \mathbf{p}_{surf}$ to $\mathbf{p}_{surf}$.

\subsection{Radiation Network Architecture}
As analyzed above, we design a 3D sparse ResNet~\cite{resnet} as the encoder and transposed convolutions as the decoder to transform the surface vibration information to the scalar-valued FFAT Map. The overall architecture is illustrated in \autoref{fig:transfer_net}.

\subsubsection{Inputs \& Outputs}
The input consists of (i) features of surface voxels and (ii) vibration frequency. 
The features of each surface voxel are concatenated by three components: normalized coordinate, vibration displacement of a voxel (averaging the displacement of 8 vertices), and a binary vector (6 elements). The binary vector indicates whether each face of a voxel is exposed to the outer space. The vibration frequency is normalized by min-max scaling in Mel scale (corresponding to 20Hz - 20000Hz).

We also design the output formulation to improve the quality of predicted scalar-valued FFAT Map. Note that the overall radiation efficiencies of different modes are different ~\cite{cremer2013structure}. As mentioned in ~\citet{PAT}, some modes are thousands of times more radiative than others. Therefore, the matrix norm of the FFAT Map (represented as a matrix) corresponding to different modes is different, as shown in \autoref{fig:transfer_norm}. This large difference can lead to difficulties in network training. Therefore, we train the network to predict the matrix norm in the log scale (i.e., FFAT Map norm) and the normalized FFAT Map separately. In other words, the network is trained to predict the overall radiation efficiency and directional difference separately.

\subsubsection{Network Architecture \& Loss Function}
There are two branches in our radiation network to predict the normalized FFAT Map and the FFAT Map norm (see \autoref{fig:transfer_net}). Each branch is an encoder-decoder-based architecture. The encoders in both branches have the same architecture, which mainly consists of a 3D sparse ResNet and two fully connected layers with activation functions. The ResNet encodes features of surface voxels and the fully connected layers encode vibration frequency. These two information codes are fused by multiplication. The encoder in FFAT Map norm branch is designed with fewer parameters (similar structure, fewer layers and channels) to prevent over-fitting. The decoder in normalized FFAT Map branch mainly consists of some 2D transpose convolutions, and the decoder in FFAT Map norm branch is a fully connected layer.

The loss function used to train the network is Mean Squared Error (MSE). The MSE loss of normalized FFAT Map and the MSE loss of FAT Map norm are summed up to compute the final loss. 

\subsubsection{Post-processing}
The scalar-valued FFAT Map is reconstructed by multiplying the FFAT Map norm with the normalized FFAT Map. Note that the radiation network is trained with a fixed voxel size. Nonetheless, our approach is applicable to any voxel size because the changes in acoustic transfer function caused by variation of voxel sizes can be handled in the same manner as~\citet{fracture}. Specifically, if the geometry of an object is scaled by $\gamma$, the multipole coefficient $M_n^m$ for acoustic transfer will be scaled as $M_{n}^{m} \rightarrow \gamma ^ {-5/2} M_{n}^{m}$. Similarly, our FFAT Map is scaled as $\mathbf{\Psi} \rightarrow \gamma^ {-5/2} \mathbf{\Psi}$.
\section{Network implementation}


\subsection{Dataset}
The 3D models we use for training and evaluation come from the ABC Dataset ~\cite{abc_dataset}, which is a large CAD model dataset for geometric deep learning. We use $\sim$100K 3D models and voxelize these models at a resolution of $32\times 32\times 32$, so that the training time and GPU memory occupied are feasible.

For the vibration solver, we use the ceramic material as representative and fix the unit voxel size at 0.5cm. 
For each 3D model, we precompute the stiffness matrix and mass matrix as well as voxel-to-vertex projection matrix and vertex-to-voxel projection matrix. These four matrices of the $\sim$100K model are saved as the dataset for our vibration solver. We classify the dataset into a training set, a test set, and a validation set with a ratio of 4:1:1. 

For the radiation network solver, we fix the resolution of the FFAT Map at $64\times 32$. For each 3D model, we use the BEM solver Bempp-cl~\cite{betcke2021bempp} to compute FFAT Maps of five randomly selected vibration modes. The surface vibration vectors are normalized before the BEM solution. To handle more general cases, we assign each 3D model a random material (one each for ceramic, glass, wood, plastic, iron, polycarbonate, steel, and tin). The input information and FFAT Maps of $\sim$100K models are saved as the dataset for our radiation solver. 
We classify the dataset into a training set, a test set, and a validation set with a ratio of 4:1:1. 

\subsection{Network Training}

We implement our networks on Minkowski Engine~\cite{mink} and Pytorch. We optimize the networks using ADAM~\cite{kingma2014adam}. The learning rates of our networks are all 1$e$-4, and the learning rate is reduced to 0.5 times every 20 epochs. For the vibration solver, $k = 20$ random starting vectors of each 3D object are concatenated into a mini-batch (i.e., the batch size is 20, and one object constitutes each mini-batch). To reduce training time cost, we generate a random subset (1/20 of the full dataset) for training at each epoch. For the radiation solver, we set the batch size to 16. We train 100 epochs for both networks and finally store the network parameters with the smallest loss in the validation set.

\section{Results}
\label{sec:results}
We highlight our experimental results along with those of the standard numerical solver and other comparable methods to evaluate the accuracy of our approach. The rendered sounds are shown in the video.

\subsection{Sound Vibration Solver}

\begin{table}[h]
\caption{Performance evaluation (average across the test dataset) of Lanczos, LOBPCG, and our solver. The first 20 modes are solved for each object. 
}
\begin{tabular}{lccc}
\hline
Method &  Conv. Error & Freq. Error & Time \\ 
\hline
Lanczos (CPU)    &  1.7$e$-6   &   1.7e-6  & 15.32s  \\  
LOBPCG (GPU)  &  1.4$e$-7  &    ~0   & 2.71s \\ 
\cline{1-4} 
Lanczos (CPU) & N/A & N/A & 0.03s \\
LOBPCG (Reduced)  & 0.165   & 0.90      & 0.03s \\ 
Ours & 0.008  & 0.10  & 0.03s  \\
\hline
\end{tabular}
\label{table:EigenNet_result}
\end{table}

We compare different vibration solvers on a randomly-selected subset (100 models) from the test dataset.
Two classic numerical methods include: (i) standard Lanczos methods~\cite{lanczos1950iteration, arpack}, which is implemented in ARPACK on a CPU (Intel i7-8700k) \cite{arpack}; (ii) standard LOBPCG~\cite{efficient_LOBPCG}, which is a iterative solver and implemented using Pytorch on a GPU (Nvidia RTX 3080Ti). 
Three metrics are used to measure the performance: residual-based convergence error (left term of \autoref{eq:loss}), Mean Square Error of frequency in Mel scale (divided by the square of the Mel scale length corresponding to 20Hz-20000Hz), and the running time cost. 
The mean value of objects' first 20 modes on these metrics is evaluated for our test set.

We first show two numerical algorithms' performance on convergence in the upper part of \autoref{table:EigenNet_result}. Both Lanczos and LOBPCG can converge, while LOBPCG is much faster, and this result also confirms the conclusion of \citet{arbenz2005comparison}. 
The converged numerical results are used as the ground truth, we then evaluate the performance of our learning-based solver in the lower part of \autoref{table:EigenNet_result}. Our solver completes network inference within 0.03s to resolve the vibration  with high accuracy in terms of both convergence error and frequency error. Using the same time budget, LOBPCG (Reduced) obtains far inaccurate results with a few iterations, while Lanczos can not produce any output.

\begin{table}[t]
\caption{Performance evaluation (average across the test dataset) of Lanczos, LOBPCG, Rayleigh-Ritz (R-R), and our mixed vibration solver in terms of different error bounds ($\varepsilon$). 
}
	\begin{tabular}{lccc}
		\toprule
			Method  &  {$\varepsilon<0.01$} & {$\varepsilon<0.005$} & { $\varepsilon<0.001$}  \\
		 \midrule
Lanczos (ILU) & 10.75s & 12.13s & 13.52s\\
Lanczos (Iterative) & 9.75s & 12.78s & 14.52s\\
R-R ($k=1$, $J=20$)     & 7.82s     & 8.14s        & 9.03s\\
R-R ($k=20$, $J=1$)     & 7.56s     & 7.77s         & 8.66s\\
R-R ($k=20$, $J=20$)     & 8.05s     & 8.31s        & 9.13s\\
LOBPCG (Reduced)   &   0.20s & 0.30s & 0.50s \\ 
Our Mixed Solver  & 0.03s & 0.12s & 0.29s\\
		\bottomrule 
	\end{tabular}
\label{table:error_metric}
\end{table}

As a warm-start initialization, our learning-based solver is integrated with LOBPCG (called mixed solver), we made further validation and highlighted the performance of our mixed solver in \autoref{table:error_metric}. We compared our mixed solver with others in terms of time cost to reach different level of accuracy, i.e., different error bounds. 

Standard Lanczos generally takes most of the time to perform the LU decomposition of the stiffness matrix, and it is not suitable for iterative refinement. Therefore, we provide two generally-used alternatives: (i) replacing the LU decomposition with incomplete LU decomposition (ILU), denoted as Lanczos (ILU), wherein the expected fill ratio upper bound of ILU decomposition is fine-tuned for different error bounds; (ii) computing the matrix inverse by an iterative solver instead of direct sparse solver, denoted as Lanczos (Iterative), wherein the number of Lanczos iterations, and number of iterative solver iterations, is fine-tuned for different error bounds.

In addition to LOBPCG and Lanczos, Rayleigh-Ritz algorithm (abbr. R-R) is also used as a baseline: starting with $k$ random vectors $\mathbf{x_1},...,\mathbf{x_k}$, the standard krylov space is spanned by vectors $(\mathbf{K}^{-1}\mathbf{M})^j\mathbf{x_i}$, where $i = 1,...,k $ and $j = 1,...,J$. Then the Rayleigh-Ritz algorithm is applied in this space to solve approximate eigenvectors. As the first 20 modes should be figured out in our experiments, we set three group settings for test: $(k = 1, J = 20)$, $(k = 20, J = 1)$, and $(k = 20, J = 20)$. The Rayleigh-Ritz algorithm is GPU-accelerated except for the LU decomposition of $\mathbf{K}$.
Like our mixed solver, the Rayleigh-Ritz algorithm works as a warm-start and then is further optimized by LOBPCG.

As can be seen from \autoref{table:error_metric}, our mixed solver consistently shows superior performance over other approaches (covering various settings) using different error bounds. Notwithstanding being inferior to our approach, LOBPCG still significantly outperforms the Lanczos and Rayleigh-Ritz algorithms.
Our mixed solver's speedup over LOBPCG decreases as the tolerance tightens (to $\sim$0.001) due to more iterations of LOBPCG being required.
Based on the above results, we choose LOBPCG (a stronger baseline) to compare the performance in the following sections.

\begin{figure}[h]
\centering
 \includegraphics[trim={0.2cm 0.2cm 0.4cm 0cm},clip, width=1.0\linewidth]{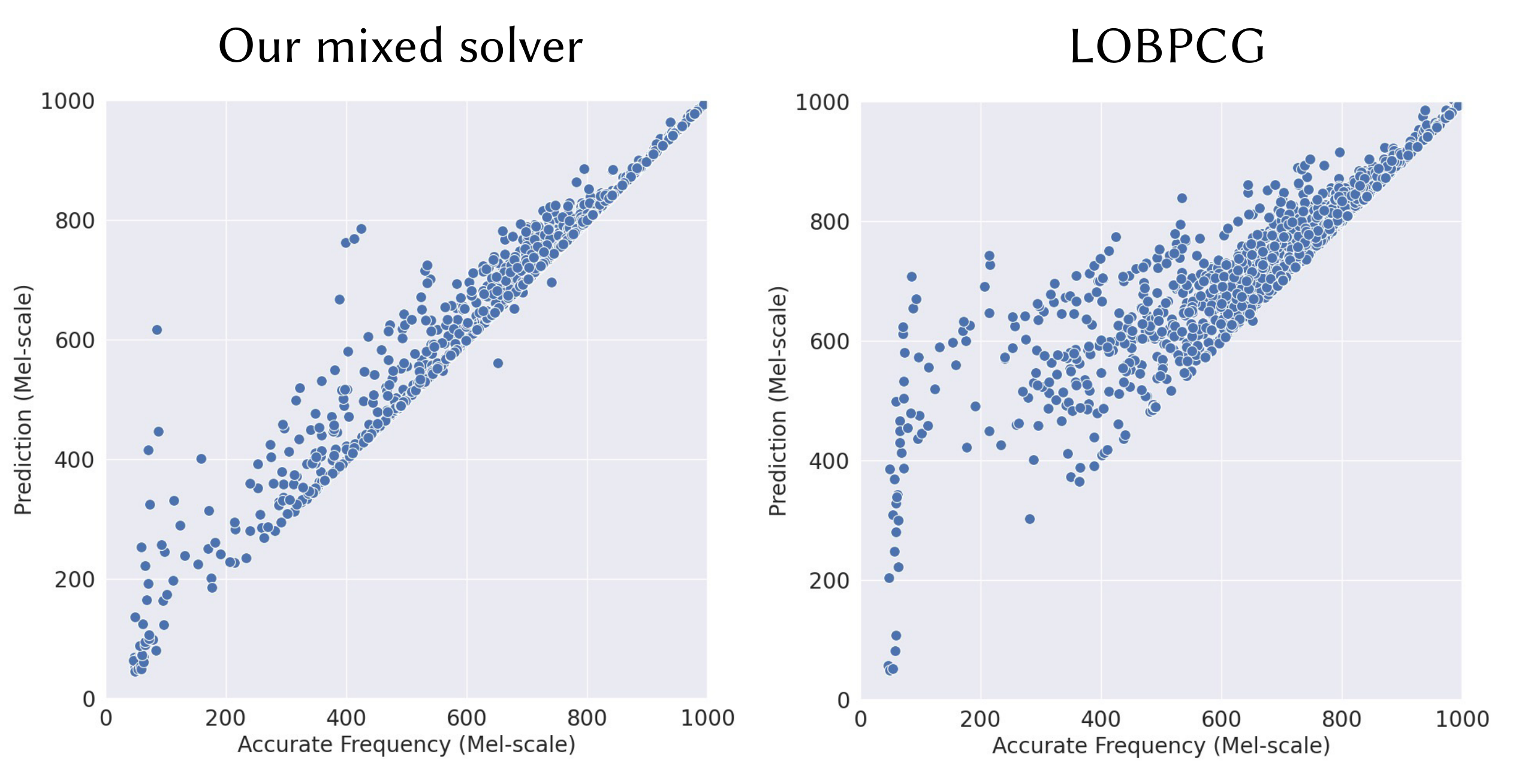}
 \caption{Visualization of mode frequencies computed by our mixed vibration solver vs. LOBPCG using same time budget (0.4s). The horizontal and vertical axes represent actual and predicted frequency in Mel scale, respectively.}
 \label{fig:freq_acc}
\end{figure}

\noindent {\bf Visualization of Results: } We draw a scatter plot to illustrate that our mixed vibration solver results in more accurate frequencies than standard LOBPCG in the same time budget, especially for low-frequency modes (see \autoref{fig:freq_acc}). Generally, a user is perceptually sensitive to the low-frequency modes with small damping coefficients than that of high-frequency. As a result, our mixed vibration solver can significantly improve the quality of sound synthesized within a limited time budget. 

\begin{figure*}[htb]
\centering
 \includegraphics[trim={0cm 0.1cm 0.5cm 0.1cm},clip,width=0.95\linewidth]{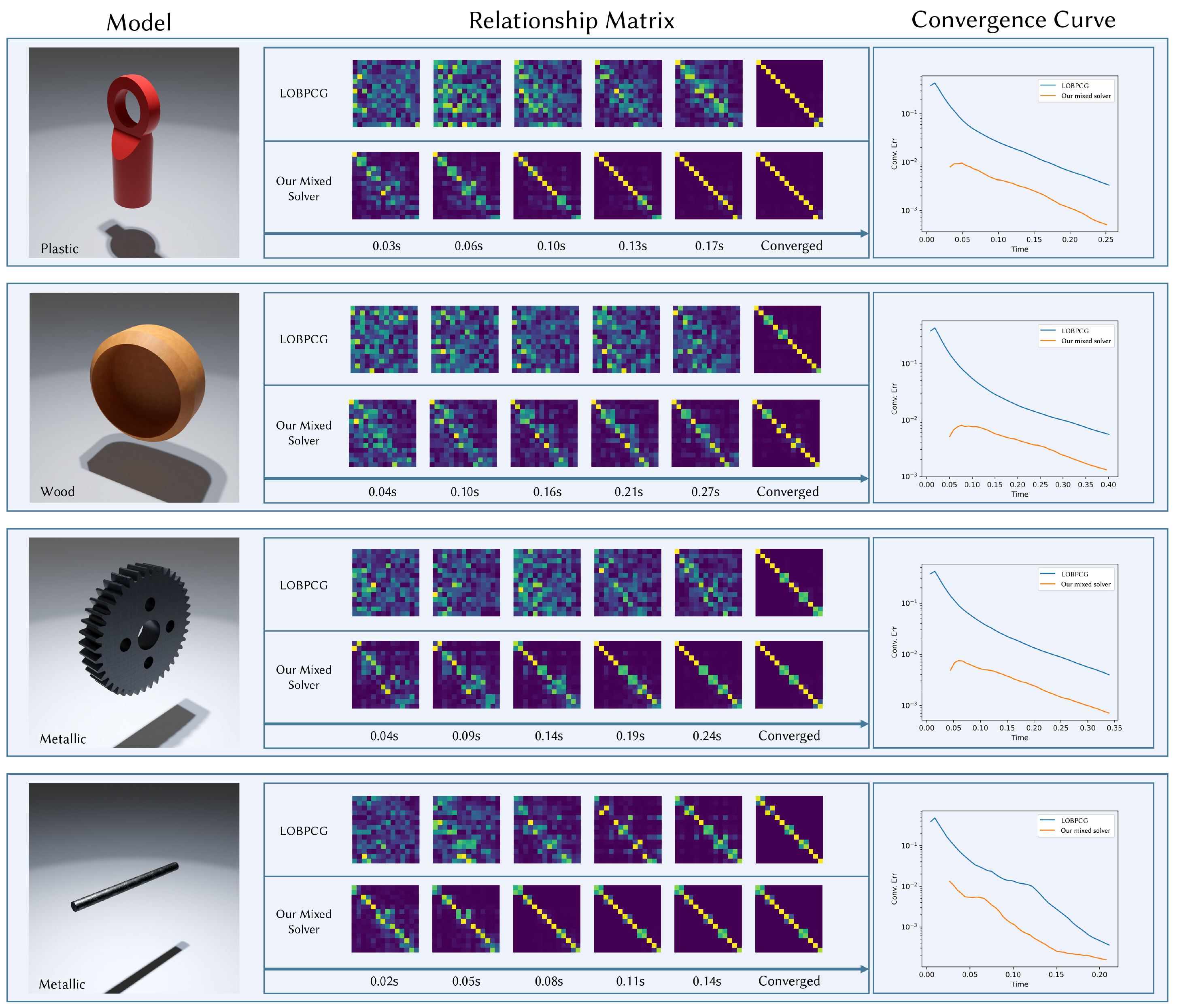}
 \caption{Performance evaluation of our mixed vibration solver and LOBPCG, visualized using relationship matrices and convergence curves of different objects with different materials. Our mixed solver results in a significant improvement in terms of convergence. 
 }
\label{fig:eigen_result}
\end{figure*}

We denote the accurate eigenvectors as $\mathbf{V}$ and the predicted eigenvectors as $\hat{\mathbf{V}}$ and plot their relationship matrix $\mathbf{V}^T \mathbf{M} \hat{\mathbf{V}}$ for different numbers of iterations. \autoref{fig:eigen_result} shows the relationship matrices within the time budget and convergence curves of our mixed vibration solver and standard LOBPCG. The results demonstrate that our mixed solver can obtain higher accuracy and result in better convergence than LOBPCG.

\begin{table}[h]
\caption{Performance evaluation (average across the test dataset) of our vibration solver and standard LOBPCG with 80 modes for each object. Our approach results in significantly lower error than LOBPCG (Reduced) within a similar time budget.}
\begin{tabular}{lccc}
\hline
Method & Conv. Error& Freq. Error      &  Time  \\ \hline
LOBPCG (GPU)             & 1.9$e$-8  &   0       & 6.52s \\
LOBPCG (Reduced)    & 0.210     &  0.33     & 0.11s \\
Ours                & 0.024     &  0.04     & 0.10s  \\

\hline
\end{tabular}
\label{table:large_mode}
\end{table}

\noindent {\bf Varying Number of Modes or Different Resolution: }
In general, a larger number of modes can enrich the sound quality. Our vibration solver also works with different numbers of modes by adjusting the number of initial vectors for each 3D object. We conduct an experiment to measure the efficiency and accuracy of our vibration solver with a larger number of modes. \autoref{table:large_mode} shows the results of our solver and standard LOBPCG with 80 modes for each object. This vibration solver is retrained with 80 initial vectors and still obtains higher accuracy than LOBPCG (Reduced) with fewer iterations using the same time budget. As compared to the performance obtained with 20 modes, as shown in \autoref{table:EigenNet_result}, one network inference to resolve 80 modes can improve the mean frequency accuracy (error 0.10 $\rightarrow$ 0.04). This may be largely because high-frequency modes are less discriminatively perceptible at Mel scale.

Rather than using a resolution of $32\times 32\times 32$, our vibration solver can also be used for higher resolution models without retraining. This is a benefit of the shift-invariance of the convolutional network and the consistency between a finite element model and a network, as highlighted in Sec. \ref{sec:consistency}. \autoref{table:highres_result} shows the results of our vibration solver and LOBPCG using the resolution of $64\times 64\times 64$ within same time budget. Our vibration solver still demonstrates superior performance over the LOBPCG, regardless of the resolution.
However, higher resolution data should take more time to complete one network inference, as shown in \autoref{table:highres_result} and \autoref{table:EigenNet_result}. 

\begin{table}[h]
\caption{Performance evaluation (average across the test dataset) of our vibration solver and LOBPCG with a resolution of $64\times 64\times 64$.  }
\begin{tabular}{lccc}
\hline
Method       & Conv. Error         & Freq. Error        &  Time  \\ \hline
LOBPCG (GPU) & 8.8$e$-7 & 0 & 12.71s \\
LOBPCG (Reduced)     & 0.134   &  1.26  & 0.12s \\
Ours       & 0.004  &  0.30  & 0.11s  \\
\hline
\end{tabular}
\label{table:highres_result}
\end{table}




\subsection{Sound Radiation Solver}
\label{sec:radiation_result}
We compare the performance of the radiation solver used in our approach with (i) BEM and (ii) a random selection method. This comparison is also made on a subset (100 models) randomly selected from the test dataset.

Note that BEM's performance is less sensitive to the number of iterations. In most cases, the construction of the boundary integrals and the evaluation of FFAT Map dominate the costs~\cite{kleinpat}. Furthermore, KleinPAT~\cite{kleinpat} is not an effective solution for our cases because the spatial range of interest is generally large ($3\times \sim 27\times$ object size), and the time-domain method requires high-resolution voxel grids, which can increase the time-stepping costs significantly. As far as we know, the only method with a comparable speed is the random selection method, i.e., a scalar-valued FFAT Map is randomly selected from the training dataset. 

We use the latest Bempp-cl library ~\cite{betcke2021bempp} as the underlying BEM solver, and various operators run on a GPU to achieve the best performance. Our radiation network runs on the same GPU.

Three metrics are used to measure the performance: MSE of the normalized FFAT Map, MSE of the log FFAT Map norm, and the running time cost. \autoref{table:AcousticNet_result} shows that our radiation solver can achieve high accuracy quickly with approximately $\sim$2200$\times$ speedup over a numerical solver to solve FFAT Maps of 20 modes, i.e., 0.04s compared to 88s of BEM. The user study in Sec. \ref{sec:user_study} also shows that the sound quality obtained by our radiation solver is close to the ground truth. Our extensive comparison (near spatial range radiation) can be found in the Appendix \ref{sec:near_radiation}.

\begin{figure*}[h]
\centering
 \includegraphics[trim={1cm 0.5cm 1.5cm 0.5cm},clip,width=0.95\linewidth]{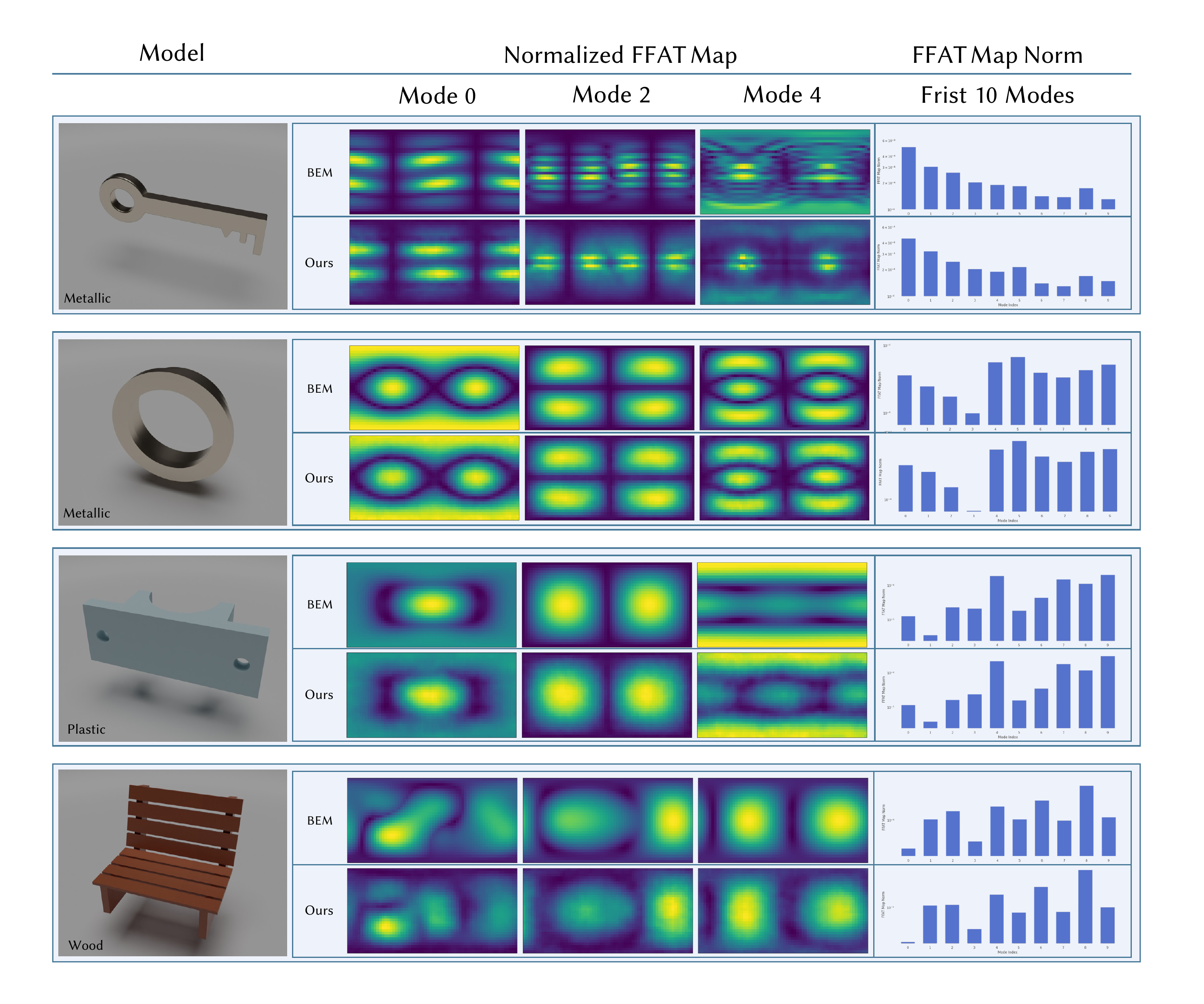}
 \caption{Results of our radiation solver and BEM show very similar structures. Normalized FFAT Map of mode 0, 2, 4 and the FFAT Map norm of the first 10 modes in log scale are plotted for comparison.
 }
\label{fig:acoustic_result}
\end{figure*}

\begin{table}[htb]
\caption{Performance evaluation (average across the test dataset) of BEM, random selection method, and our radiation solver. FFAT Maps of the first 20 modes are solved for each object.}
\begin{tabular}{lccc}
\hline
\multirow{2}{*}{Method} & \multirow{2}{*}{\begin{tabular}[c]{@{}c@{}}Normalized \\ FFAT Map MSE\end{tabular}} & \multirow{2}{*}{Log Norm MSE} & \multirow{2}{*}{Time} \\
 &  &  &  \\ \hline
BEM & 0 & 0 & 88s \\
Random Selection & 0.63 & 4.76 & 0s \\
Ours & 0.06 & 0.07 & 0.04s \\ \hline
\end{tabular}
\label{table:AcousticNet_result}
\end{table}

\noindent {\bf Visualization of Results: }
We similarly visualize the normalized FFAT Map as KleinPAT ~\cite{kleinpat} and the distribution of FFAT Map norm (in log scale) for several objects, to compare our radiation network solver with BEM. Our results are similar to the results of BEM (see \autoref{fig:acoustic_result}). Overall, our radiation solver works better at low frequencies than at high frequencies. 
Furthermore, we observe that the FFAT Map Norm distribution can significantly affect the pitch of a sound, which humans are sensitive to, so its accuracy is critical.
Our radiation solver can accurately predict the FFAT Map Norm distribution.

\subsection{Extensive Cases}
Our learning-based sound synthesis approach enables the efficient synthesis of different desirable sound effects. We demonstrate some results with comparisons and applications. All the animations are generated using the physically-based simulator Pybullet~\cite{pybullet} except the Jumping Jelly, which is an artistically designed animation. The first three 3D models are used in \citet{kleinpat}, and the Tin Bell model is from~\citet{PAT}. Other models are designed using Blender software.  

\begin{figure*}
     \centering
    \begin{subfigure}[b]{0.24\textwidth}
         \centering
         \includegraphics[width=\textwidth]{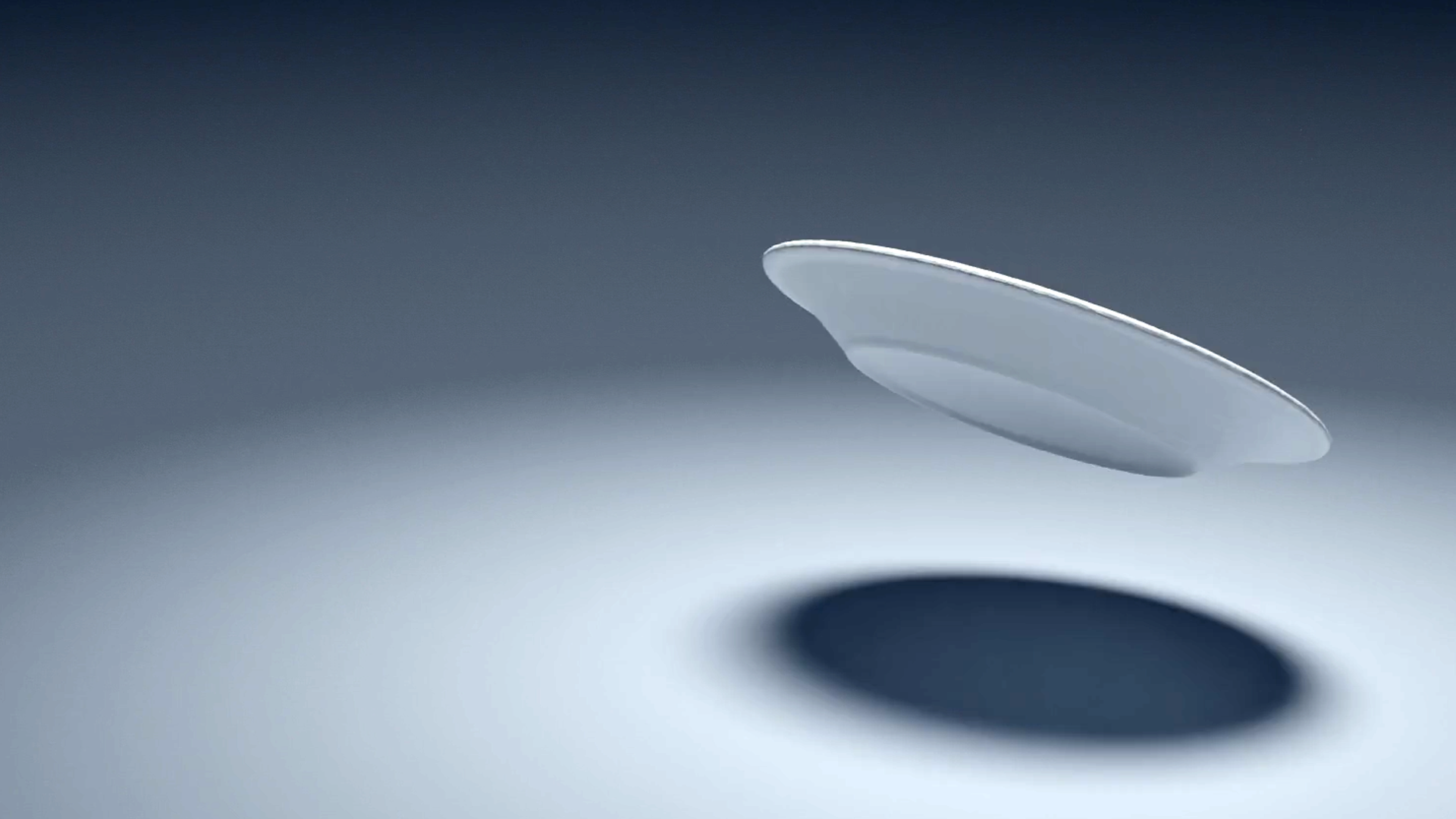}
         \caption{Ceramic Plate}
         \label{fig:CeramicPlate}
     \end{subfigure}
    \begin{subfigure}[b]{0.24\textwidth}
         \centering
         \includegraphics[width=\textwidth]{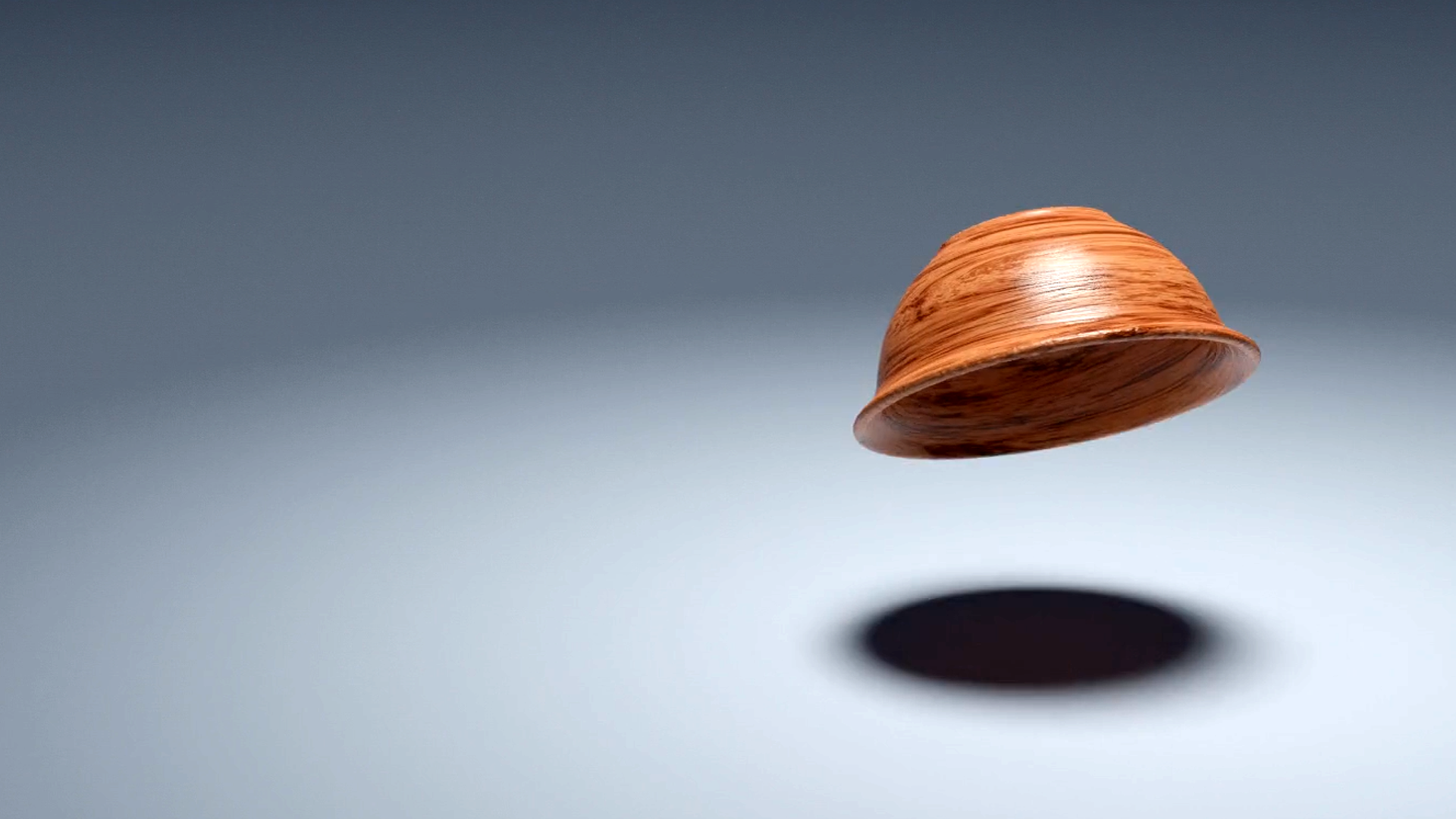}
         \caption{Wood Bowl }
         \label{fig:WoodBowl}
     \end{subfigure}
    \begin{subfigure}[b]{0.24\textwidth}
         \centering
         \includegraphics[width=\textwidth]{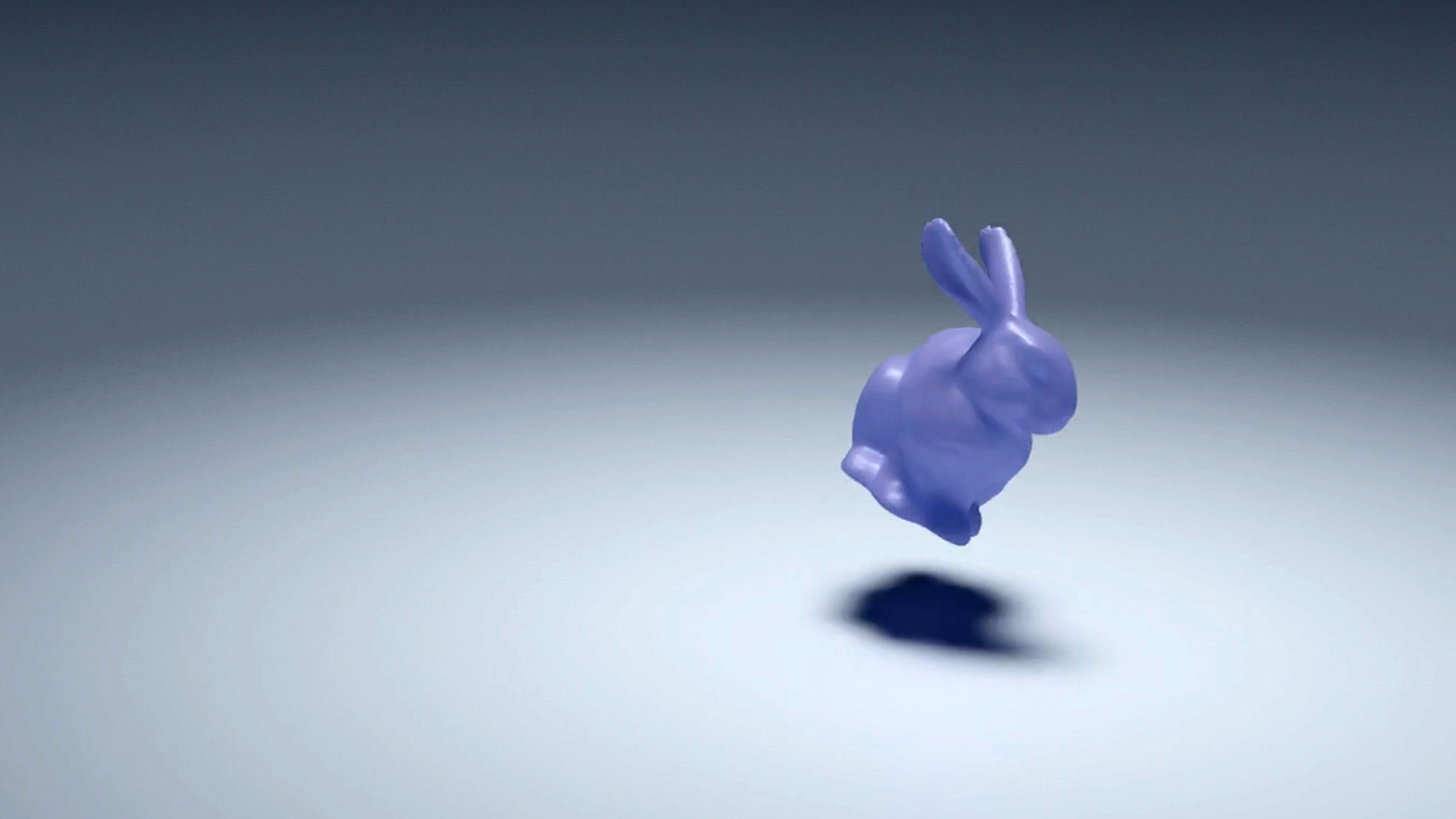}
         \caption{Plastic Bunny}
         \label{fig:PlasticBunny}
     \end{subfigure}
    \begin{subfigure}[b]{0.24\textwidth}
         \centering
         \includegraphics[width=\textwidth]{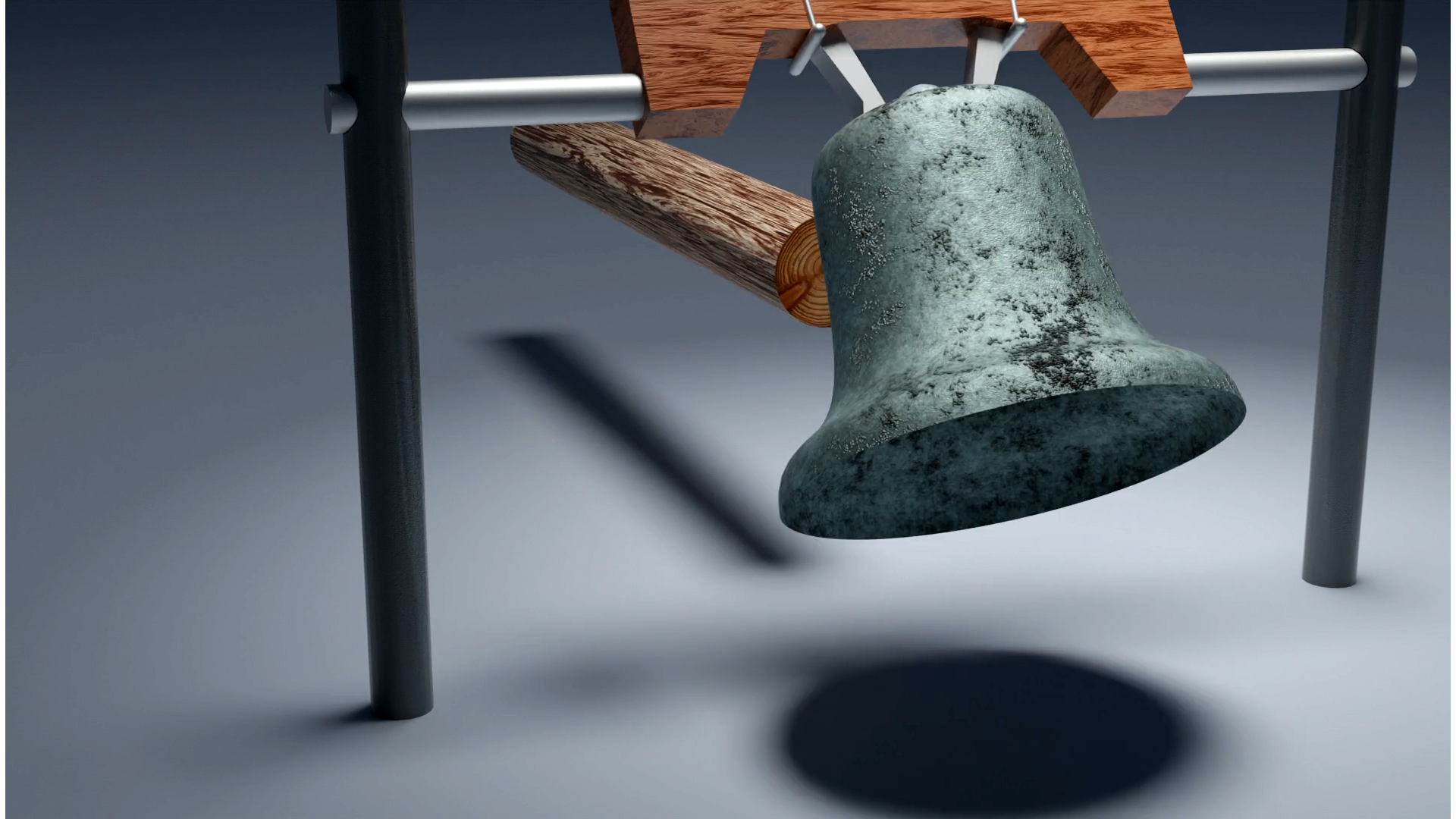}
         \caption{Tin Bell}
         \label{fig:TinBell}
     \end{subfigure}
    \begin{subfigure}[b]{0.24\textwidth}
         \centering
         \includegraphics[width=\textwidth]{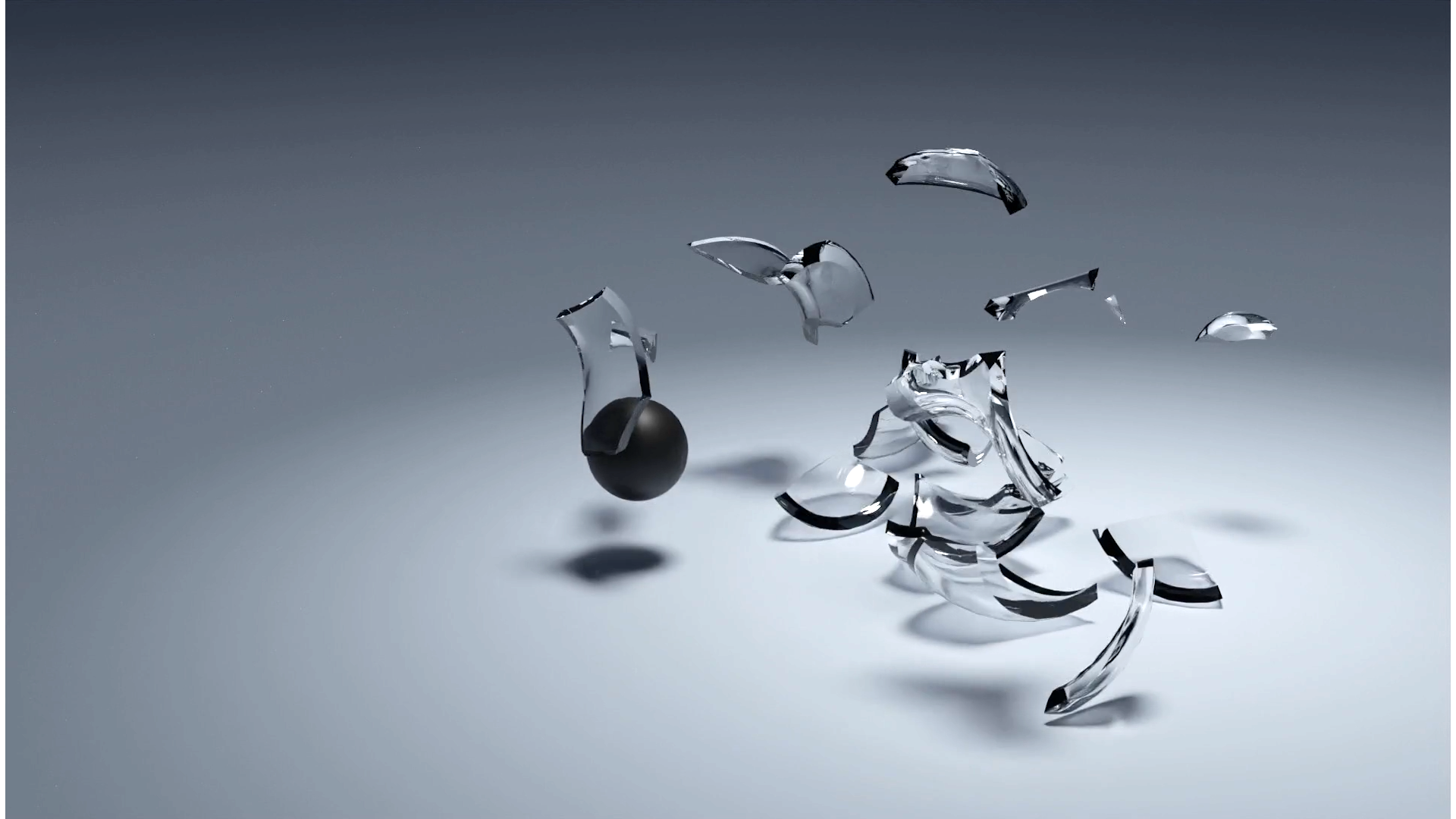}
         \caption{Glass Vase Fracture}
         \label{fig:GlassVaseFracture}
     \end{subfigure}
    \begin{subfigure}[b]{0.24\textwidth}
         \centering
         \includegraphics[width=\textwidth]{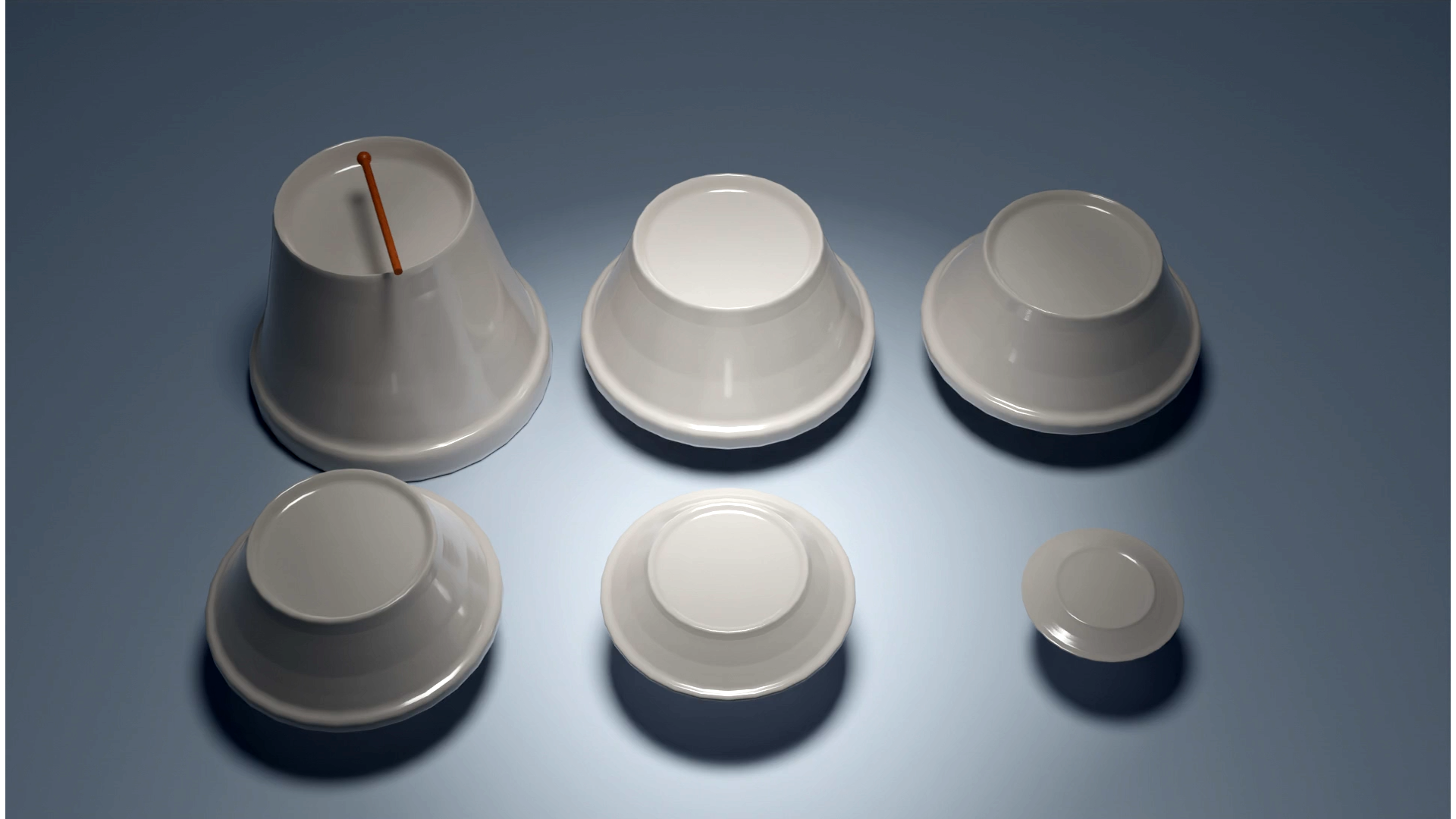}
         \caption{Little Star }
         \label{fig:LittleStar}
     \end{subfigure}
    \begin{subfigure}[b]{0.24\textwidth}
         \centering
         \includegraphics[width=\textwidth]{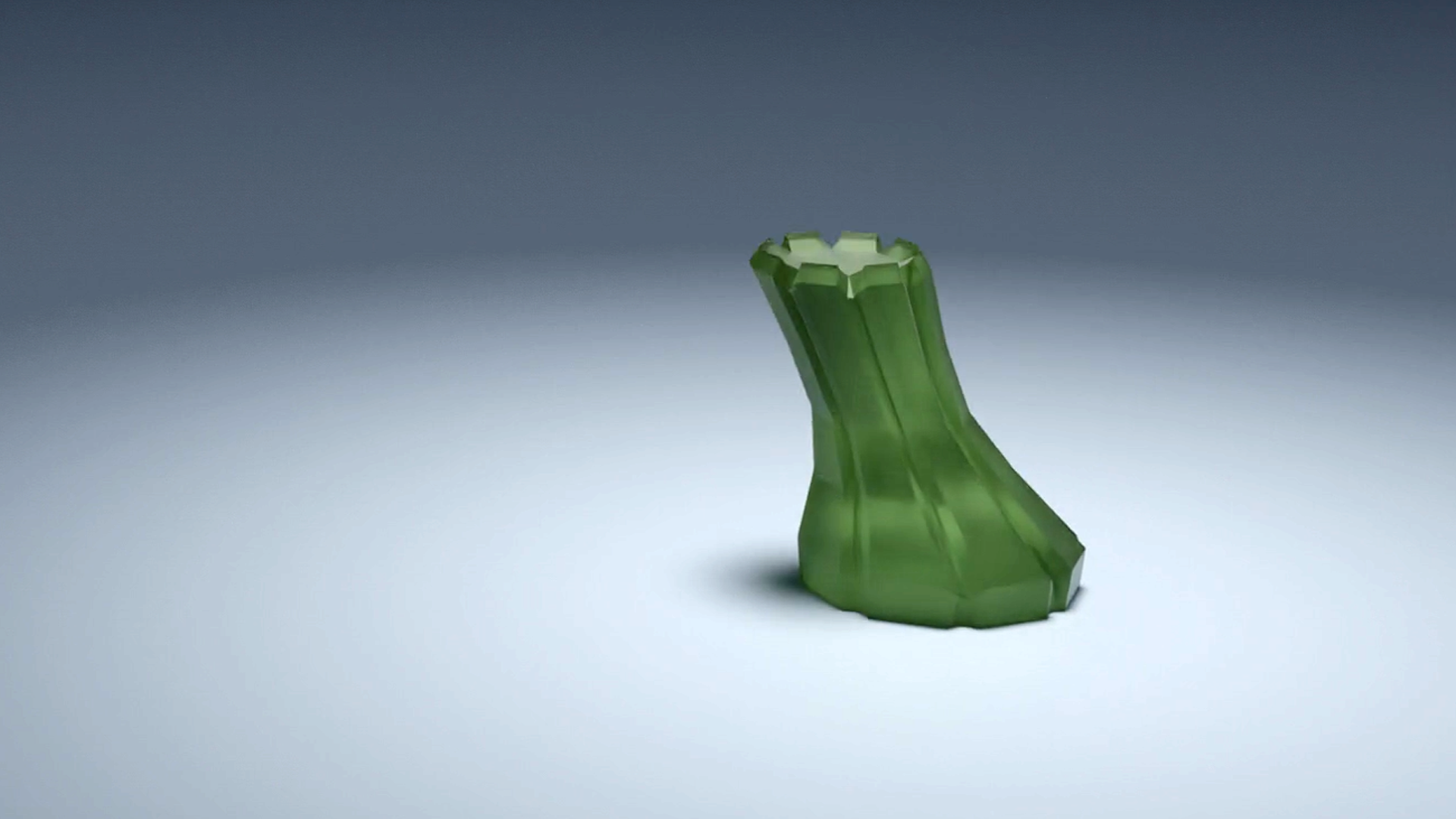}
         \caption{Jumping Jelly}
         \label{fig:Jelly}
     \end{subfigure}
    \begin{subfigure}[b]{0.24\textwidth}
         \centering
         \includegraphics[width=\textwidth]{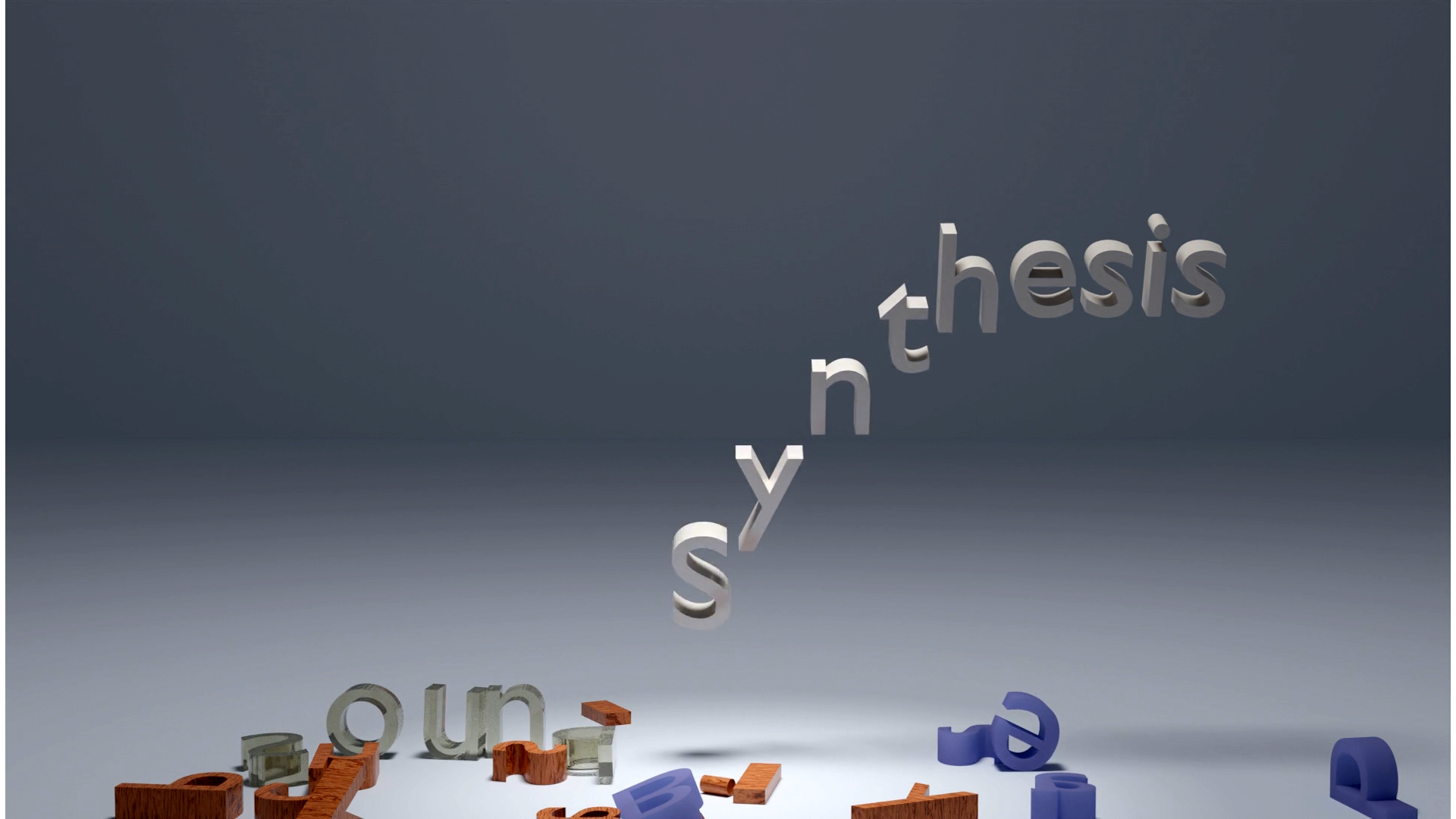}
         \caption{Letters}
         \label{fig:Letters}
     \end{subfigure}
        \caption{(\textbf{a}, \textbf{b}, \textbf{c}, \textbf{d}, \textbf{e}) We compare our approach with other baselines on 3 drop scenes, a ceramic plate (20 modes), a wood bowl (40 modes), and a plastic bunny (20 modes), as well as a ringing bell scene (80 modes) and a glass fracture scene (17 fractures, each 20 modes). (\textbf{f}) We play the melody "Little Star" by tapping bowls of different shapes and sizes. The vibration and radiation data after editing the model are solved by our approach. (\textbf{g}) Our approach can generate artistic sound effects with time-varying frequencies and acoustic transfer in a soft body animation. (\textbf{h}) Our approach enables fast sound synthesis for scenes with multiple moving objects, where we use 20 modes for each object. }
        \label{fig:demo}
\end{figure*}

\noindent {\bf Fast Editable Shapes: } 
To achieve certain sound characteristics, the user might rely on a trial-and-error approach to tune the material parameters~\cite{Interactive-AT}, and a fast sound synthesis method can shorten the tuning cycle. Our learning-based sound synthesis approach can not only shorten the material tuning cycle but also the model shape tuning cycle. Each adjust-synthesis cycle time is less than 2 seconds in the "Little Star" benchmark (see \autoref{fig:LittleStar}). 

\noindent {\bf Fast Precomputation:}
Precomputation will needs more time to handle more objects. Our learning-based sound synthesis approach can significantly accelerate the computation. It takes less than one minute to precompute all the sound vibration and radiation data by our approach for a scene with 36 letter objects (see \autoref{fig:Letters}). On the other hand, numerical solvers (standard LOBPCG + BEM) takes about one hour for sound synthesis methods.

\noindent {\bf Artistic Effect Design: } 
In \citet{Interactive-AT}, user-specific non-physical time-varying frequencies are used to generate interesting artistic effects in soft-body animation. By approximating the soft body model in each frame as a different object, our learning-based sound synthesis approach can quickly generate interesting sound effects synchronized with the animations (see \autoref{fig:Jelly}).

\section{Perceptual Evaluation}
\label{sec:user_study}

We conducted a preliminary user study to evaluate the sound quality generated by our learning-based sound synthesis approach. Our goal was to verify that our approach generates plausible sounds similar to the reference sounds. A total of 106 subjects (each with normal hearing) were enrolled in this experiment, and each subject used a pair of headphones for sound playback. Each user performed two tests (vibration test and radiation test):

    \textbf{Vibration test:} Five videos, as shown in \textbf{a}-\textbf{e} of \autoref{fig:demo}, cover a variety of materials, including a ceramic plate (20 modes), a wood bowl (40 modes), a plastic bunny (20 modes), a tin bell (80 modes), and a glass vase (17 fractures, 20 modes each fracture). In each video clip, three sound vibration solvers are applied to generate sound consistent with the animation: (i) LOBPCG (ground truth as the \emph{reference}), (ii) our mixed vibration solver, and (iii) LOBPCG (reduced) using the same time budget, with (ii) as the \emph{baseline}.
    During this test, the reference sound (i) was played first, then (ii) and (iii) were played in a random order, and a counter balance was used to control the order effect. The subjects were asked to measure the similarity between (ii) and (i) as well as between (iii) and (i) using a 7-point Likert scale ranging from 1 (no similarity at all) to 7 (no difference at all). Note that the subject did not have any prior knowledge about which one from (ii) and (iii) was produced by our approach. 
    
   \textbf{Radiation test:} Sound benchmarks from the vibration test were also used to evaluate the radiation solver. In each video clip, three sound radiation solvers were applied to generate sound consistent with the animation: (i) BEM (ground truth as the \emph{reference}), (ii) our radiation solver, and (iii) the random selection method described in Sec.~\ref{sec:radiation_result} as the \emph{baseline}.
   During this test, the reference sound (i) was played first, then (ii) and (iii) were played at random, and a counter balance was used to control the order effect. The subjects were asked to measure the similarity between (ii) and (i) as well as (iii) and (i) using the same Likert scale used in the vibration test. Note that the subject did not have any prior knowledge about which one from (ii) and (iii) was produced by our approach.

\begin{table}[]
\caption{Mean value with standard deviation obtained from the \emph{vibration test} and the \emph{radiation test}, respectively. Each scene and test had all 106 subjects. Our approach shows very high fidelity to the ground truth and significantly outperforms the baseline in both the vibration and radiation tests.}
\begin{tabular}{ccccc}
\hline
\multirow{2}{*}{Scene} & \multicolumn{2}{c}{Vibration} & \multicolumn{2}{c}{Radiation} \\ \cline{2-5} 
 & Ours & Baseline & Ours & Baseline \\ \hline
\textbf{a} & 6.55 $\pm$ 0.69 & 1.95 $\pm$ 1.11 & 6.65 $\pm$ 0.66 & 2.53 $\pm$ 1.49\\
\textbf{b} & 6.49 $\pm$ 0.73 & 2.53 $\pm$ 1.42 & 6.15 $\pm$ 1.01 & 5.02 $\pm$ 1.71\\
\textbf{c} & 6.30 $\pm$ 0.87 & 4.45 $\pm$ 1.75 & 6.58 $\pm$ 0.74 & 3.75 $\pm$ 1.68\\
\textbf{d} & 6.46 $\pm$ 0.79 & 3.53 $\pm$ 1.85 & 5.77 $\pm$ 1.15 & 3.45 $\pm$ 1.80\\
\textbf{e} & 5.67 $\pm$ 1.45 & 4.03 $\pm$ 1.73 & 6.39 $\pm$ 0.84 & 5.03 $\pm$ 1.61\\
ALL & 6.29 $\pm$ 1.00 & 3.30 $\pm$ 1.85 & 6.31 $\pm$ 0.95 & 3.96 $\pm$ 1.92\\ \hline
\end{tabular}
\label{tab:userStudy1}
\end{table}

The descriptive statistical analysis for different methods and scenes is shown in \autoref{tab:userStudy1}. We show the mean value with standard deviation of similarity obtained from the \emph{vibration test} and the \emph{radiation test}. Our approach shows very high fidelity to the ground truth (averaging 6.29 for sound vibration and 6.31 for sound radiation, respectively) and significantly outperforms the baseline in both the vibration and radiation tests.

As to the performance differences across all five scenes, our vibration solver obtain a significantly lower score on scene \textbf{e} (glass fracture) than other scenes ($p$ $<$ .001 when compared with scenes \textbf{a} - \textbf{d}), and our radiation solver obtain a lower score on scene \textbf{d} (tin bell) than others scenes ($p$ = .055 when compared with scene \textbf{b}, and $p$ $<$ .001 when compared with scenes \textbf{a}, \textbf{c}, \textbf{e}).

To further evaluate the difference between our approach and the baselines, we employed two-way repeated measures ANOVAs (Analysis of Variance) with the within-subjects factor \textbf{method} (ours, baselines) and \textbf{scene} (\textbf{a} - \textbf{e} of \autoref{fig:demo}) for  vibration and radiation individually.  

\begin{table}[h]
\centering
\caption{Simple effect of methods (ours vs. baseline) on scores in each test scene (Bonferroni correction applied). Our approach achieves significantly higher scores than baselines in both the vibration and radiation tests.}
\begin{tabular}{lcccc}
\hline
\multirow{2}{*}{Scene} & \multicolumn{2}{c}{Vibration} & \multicolumn{2}{c}{Radiation} \\ \cline{2-5} 
 & Mean Diff. & $p$-value & Mean Diff. & $p$-value \\ \hline
\textbf{a} & 4.594 & $<$.001 & 4.123 & $<$.001 \\
\textbf{b} & 3.962 & $<$.001 & 1.132 & $<$.001 \\
\textbf{c} & 1.849 & $<$.001 & 2.821 & $<$.001 \\
\textbf{d} & 2.934 & $<$.001 & 2.321 & $<$.001 \\
\textbf{e} & 1.642 & $<$.001 & 1.358 & $<$.001 \\ \hline
\end{tabular}
\label{posthocmethod}
\end{table}
{\bf Vibration: }
There is a significant main effect of \textbf{method} ($F(1, 105)$ = 442.432, $p$ $<$ .001), a significant main effect of \textbf{scene} ($F(4, 420)$ = 42.154, $p$ $<$ .001), and a significant interaction between \textbf{method} and \textbf{scene} ($F(4, 420)$ = 101.731, $p$ $<$ .001) on participants' scores. As shown in the left side of ~\autoref{posthocmethod}, Bonferroni-adjusted comparisons indicate that our vibration solver outperforms the baseline, i.e., LOBPCG (reduced), significantly in all five scenes. 

\textbf{Radiation}: There is a significant main effect of \textbf{method} ($F(1, 105)$ = 365.844, $p$ $<$ .001), a significant main effect of \textbf{scene} ($F(4, 420)$ = 71.225, $p$ $<$ .001), and a significant interaction between \textbf{method} and \textbf{scene} ($F(4, 420)$ = 68.180, $p$ $<$ .001) on participants' scores. As shown in the right side of ~\autoref{posthocmethod}, Bonferroni-adjusted comparisons indicate that our radiation solver outperforms the baseline, i.e. the random selection method, significantly in all five scenes.  


Overall, the results show that the sounds synthesized by our approach are much closer to the ground truth than the sounds synthesized by the baselines.

\section{conclusion, Limitations, and Future RESEARCH}

We present a novel learning-based sound synthesis approach. We design our vibration solver based on the connection between a 3D sparse U-Net and the numerical solver based on matrix computations. Our vibration solver can compute approximate eigenvectors and eigenvalues quickly. The accuracy of the results is further optimized by an optional LOBPCG solver (mixed vibration solver).
We design our radiation solver as an end-to-end network based on the connection between convolution neural networks and BEM. We evaluate the accuracy and speed on many benchmarks and highlight the benefits of our method in terms of performance.
Our approach has some limitations. The hyperparameters used in our network have not been fine-tuned. The vibration solver is not much faster than the standard LOBPCG, especially under highly tight-tolerance conditions. Despite good performance on some objects, our radiation solver does not predict with good accuracy in high-frequency. 

There are many avenues for future work. In addition to overcoming these limitations, developing a neural network that can be equivalent to the traditional iterative solution with fast convergence is an interesting area of research. We need to evaluate our approach on other benchmarks and complex scenarios, e.g., multi-contact of multiple objects~\cite{zhang2015quadratic} or frictional scenarios. 
A better predictor in radiation for high frequencies can further improve the quality of sound acoustic transfer. Finally, we would expect to integrate and evaluate our method with learning-based sound propagation algorithms~\cite{https://doi.org/10.48550/arxiv.2204.01787,https://doi.org/10.48550/arxiv.2110.04057} and use these methods for interactive applications, including games and VR.

\begin{acks}
We thank Mr. Xiang Gu for his helpful suggestions on the user study.
\end{acks}

\bibliographystyle{ACM-Reference-Format}
\bibliography{NeuralSound}

\appendix

\section{appendix}

\subsection{Analogy Between Network and Matrix Computation}
\label{sec:valid_network}

\subsubsection{Assembled Matrix and 3D Sparse Convolution}

Assuming a finite element model with $N$ hexahedrons and $M$ vertices, an assembled matrix (e.g. mass matrix or stiffness matrix) $\mathbf{A} \in \mathbb{R}^{3M \times 3M}$ is built by assembling the element matrix $\mathbf{A_e} \in \mathbb{R} ^ {24 \times 24}$ for all hexahedrons. 
A vector $\mathbf{x} \in \mathbb{R}^{3M}$ represents the displacements of all vertices (in $x, y ,z$ directions). A equivalent form $\mathbf{x'} \in \mathbb{R}^{24N}$ represents the displacements of all hexahedrons, where the displacement of each hexahedron consists of the displacements of all its vertices. The matrix-vector multiplication $\mathbf{Ax}$ has an equivalent form $\mathbf{(Ax)'}$, which satisfies:
\begin{equation}
 \mathbf{(Ax)'}_{\mathbf{u}}=\sum_{\mathbf{i} \in \mathcal{N}\left(\mathbf{u}\right)} \mathbf{W}_{\mathbf{u},\mathbf{i}}\mathbf{x'}_{\mathbf{u}+\mathbf{i}} \ ,
 \label{eq:assembled_matrix_convolution}
\end{equation}
where $\mathbf{x'}_\mathbf{u}, \mathbf{(Ax)'}_\mathbf{u}$ represents the displacement of the hexahedron at $\mathbf{u}$ before and after multiplication, $\mathbf{u} \in \mathbb{Z}^{3}$ is the 3D coordinate of a hexahedron, $\mathcal{N}\left(\mathbf{u}\right)=\{(x,y,z)|-1\leq x,y,z \leq 1\ , x,y,z\in \mathbb{Z}\}$ is the set of coordinate offsets from the current hexahedron to the neighboring hexahedrons (including itself), and $\mathbf{W}_{\mathbf{u},\mathbf{i}} \in \mathbb{R}^{24\times24}$ is the transform matrix for the hexahedron at $u$ and the coordinate offset $\mathbf{i}$. When $\mathbf{W}_{\mathbf{u}, \mathbf{i}}$ is independent of $\mathbf{u}$, \autoref{eq:assembled_matrix_convolution} can also be regarded as the definition of 3D sparse convolution~\cite{mink}. Assuming the element matrix $\mathbf{A_e}$ is fixed, $\mathbf{W}_{\mathbf{u}, \mathbf{i}}$ is independent of $\mathbf{u}$ as:
\begin{equation}
    \mathbf{W}_{\mathbf{i}}(j,k) = 
    \begin{cases}
        0& Vertex(j) \not\in Voxel(\mathbf{u} + \mathbf{i})\\
        \mathbf{A_e}(j',k) & Vertex(j) \in Voxel(\mathbf{u} + \mathbf{i})
    \end{cases}
\end{equation}
where $j$ is the index of a vertex $\mathbf{v}_j$ in the hexahedron at $\mathbf{u}$, $j'$ is the index of the vertex $\mathbf{v}_j'$ in the neighbor hexahedron $\mathbf{u} + \mathbf{i}$. $\mathbf{v}_j$ and $\mathbf{v}_j'$ coincide in coordinates and $k$ is the index of any vertex in a hexahedron. Therefore, the matrix-vector multiplication corresponds to a $3 \times 3 \times 3$ sparse convolution. 

We also conduct an experiment to validate that a 3$\times$3$\times$3 sparse convolution is equivalent to an assembled matrix. We train a network $\mathbf{g}$ (with one 3$\times$3$\times$3 3D sparse convolution, see bottom right of \autoref{fig:eigen_net}) with parameters $\mathbf{\theta}$ on a dataset with $N$ objects by reducing the mean relative error:
\begin{equation}
\hat{\mathbf{\theta}}=\underset{\mathbf{\theta}}{\operatorname{argmin}} \frac{1}{N} \sum_{i=1}^{N} 
   \frac{||g(\mathbf{x}_i ; \boldsymbol{\theta}) - \mathbf{A}_i\mathbf{x}_i||_2}{||\mathbf{A}_i||_2 ||\mathbf{x}_i||_2} \ ,
   \label{eq:validation_loss}
\end{equation}
where $\mathbf{A}_i, \mathbf{x}_i$ are the assembled matrix and the random vector of the $i$th object, respectively. For each type of assembled matrix (e.g., stiffness matrix for ceramic objects), we retrain the network and list the mean relative error on the test set after 10 epochs in \autoref{tab:appendix1}. The result shows that this sparse convolution can be trained to represent an assembled matrix with high accuracy and validates our analysis in Sec. \ref{sec:consistency}.

\begin{table}[hb]
\caption{Mean relative error between an assembled matrix and a 3$\times$3$\times$3 sparse convolution in test set. We choose the assembled matrix as mass and stiffness matrices with different materials.}
\begin{tabular}{l|llll}
\hline
            & ceramic & steel & plastic & glass \\ \hline
stiffness   & 7e-4    & 5e-4  & 6e-4  & 6e-4    \\
mass        & 6e-4    & 7e-4  & 6e-4  & 7e-4       \\ \hline
\end{tabular}
\label{tab:appendix1}
\end{table}

\subsubsection{Inverse Assembled Matrix and Sparse U-Net}
\begin{figure}[hbt]
\centering
 \includegraphics[trim={0.5cm 0cm 1cm 1cm},clip,width=0.8\linewidth]{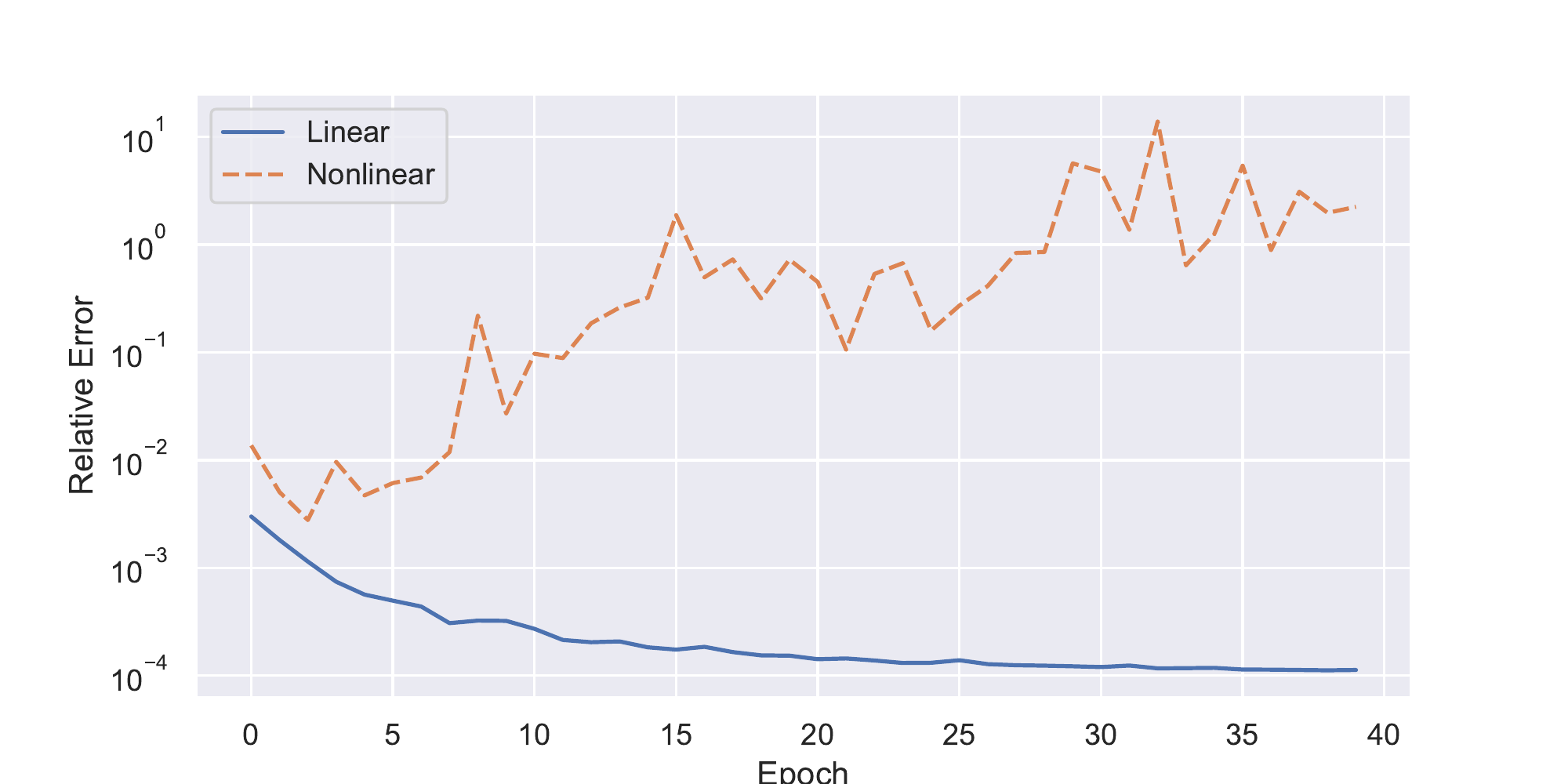}
 \caption{Mean relative error of 3D sparse linear U-Net and nonlinear U-Net in test set. 
 }
\label{fig:linear}
\end{figure}

We conduct an experiment to validate that a 3D sparse U-Net can approximate the inverse of an assembled matrix. For a U-Net $g$ with parameters $\mathbf{\theta}$, we train it on a dataset with $N$ objects by reducing the mean relative error:
\begin{equation}
\hat{\mathbf{\theta}}=\underset{\mathbf{\theta}}{\operatorname{argmin}} \frac{1}{N} \sum_{i=1}^{N} \frac{||\mathbf{A}_ig(\mathbf{x}_i ; \boldsymbol{\theta})-\mathbf{x}_i||_2}{||\mathbf{x}_i||_2} \ .
   \label{eq:validation_loss2}
\end{equation} Where $\mathbf{A}_i, \mathbf{x}_i$ are the assembled matrix and the random vector of the $i$th object.
We train two types of U-Net respectively, including a standard 3D sparse U-Net (with Relu activation) and a 3D sparse linear U-Net (without nonlinear activation). We plot their mean relative error in the test set for first 40 epochs, as shown in \autoref{fig:linear}. The standard nonlinear U-Net cannot converge, while the linear U-Net converges to a low error and validates our theoretical analysis in Sec. \ref{sec:consistency}.

\begin{figure}[hbt]
\centering
 \includegraphics[trim={0.5cm 0cm 1cm 1cm},clip,width=0.8\linewidth]{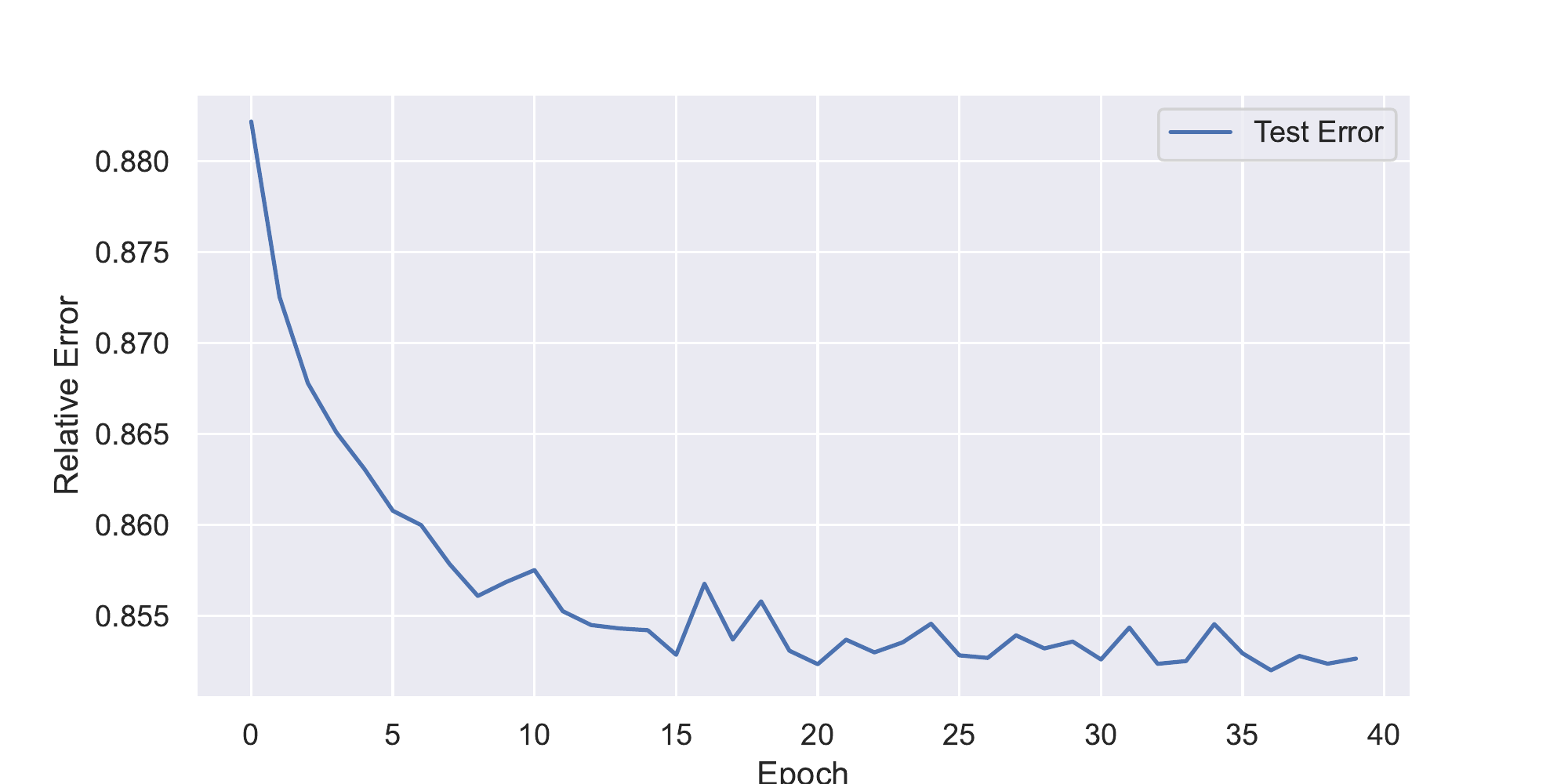}
 \caption{Mean relative error in test set of 3D sparse linear U-Net trained without the Rayleigh-Ritz method.
 }
\label{fig:ritz}
\end{figure}

\subsection{Solving Eigenvectors in Krylov Subspace}
\label{example_solve_eigenvectors}
In the generalized eigenvalue decomposition
$\mathbf{K U}= \mathbf{M} \mathbf{U} \mathbf{\Lambda}$, the $i$th eigenvector $\mathbf{u}_i$ and eigenvalue $\lambda_i$ ($\lambda_i$ > 0) satisfy
$\mathbf{K}\mathbf{u}_i = \lambda_i \mathbf{M} \mathbf{u}_i \ .$
The initial vector, $\mathbf{b}$, can be written as a linear combination of all the eigenvectors (assuming there are $n$ eigenvectors):
\begin{equation}
    \mathbf{b} = \sum_{i = 1}^n c_i\mathbf{u}_i \ ,
\end{equation}
where $c_i$ is the linear coefficient. Multiplying the matrix $\mathbf{K}^{-1} \mathbf{M}$ to $\mathbf{b}$, then:
\begin{equation}
\begin{aligned}
 \mathbf{K} ^{-1} \mathbf{M} \mathbf{b} =  \sum_{i = 1}^n c_i \mathbf{K}^{-1} \mathbf{M} \mathbf{u}_i 
= \sum_{i = 1}^n \frac{c_i}{\lambda_i}\mathbf{u}_i \ .
\end{aligned}
\label{eq:iteration}
\end{equation}
In modal analysis, the first $k$ smallest eigenvalues and corresponding eigenvectors with respect to the human audible perception range (usually 20HZ $\sim$ 20000HZ) are used. From \autoref{eq:iteration}, the weight of the eigenvector $\mathbf{u}_i$ is re-scaled from $c_i$ to $\frac{c_i}{\lambda_i}$. Therefore, eigenvectors with larger eigenvalues will have smaller weights, and eigenvectors with smaller eigenvalues will dominate the result of this linear combination. Suppose there are $k$ items from multiplying the matrix $\mathbf{K} ^{-1} \mathbf{M}$ to different initial vectors: $\mathbf{K} ^{-1} \mathbf{M} \mathbf{b}_0, \mathbf{K} ^{-1} \mathbf{M} \mathbf{b}_1, ..., \mathbf{K} ^{-1} \mathbf{M} \mathbf{b}_k$, then the Rayleigh–Ritz method~\cite{lobpcg, ritz1, ritz2} can be used to compute the first $k$ smallest approximate eigenvalues and corresponding eigenvectors in the subspace  $S = span\{ \mathbf{K} ^{-1} \mathbf{M} \mathbf{b}_i\}$, $i = 1,2,...,k$. 

In addition to $\mathbf{K}^{-1} \mathbf{M}$, other matrices (e.g., $\left(\mathbf{K} - \hat \lambda \mathbf{M}\right)^{-1}\mathbf{M}$ for any $\hat \lambda$) can also generate a subspace to compute the approximate eigenvectors in a similar manner.

\subsection{Effectiveness of Using Rayleigh–Ritz Method}
\label{sec:valid_ritz}
We add an experiment to validate the function of Rayleigh–Ritz method by taking it out as an ablation study. Specifically, we precompute the first 20 eigenvectors of $N$ object in our dataset. Then we train a 3D linear sparse U-Net $g$ with parameters $\mathbf{\theta}$ on this dataset by reducing the mean relative error between the accurate eigenvectors (ground-truth) and the predicted eigenvectors:
\begin{equation}
\hat{\mathbf{\theta}}=\underset{\mathbf{\theta}}
{\operatorname{argmin}}\frac{1}{N} \sum_{i = 1}^{N}
   \frac{1}{20} \sum_{j = 1}^{20} \frac{||g(\mathbf{x}_{i,j} ; \boldsymbol{\theta})- \mathbf{y}_{i,j}||_2}{||\mathbf{y}_{i,j}||_2} \ ,
 \label{eq:validation_loss_3}
\end{equation}
where $\mathbf{y}_{i,j}, \mathbf{x}_{i,j}$ are the $j$th ground-truth eigenvector and the $j$th random vector of the $i$th object. We plot the mean relative error in test set for first 100 epochs in \autoref{fig:ritz}. Without the Rayleigh–Ritz method, the 3D U-Net can only converge to a result (error $\approx$ 0.85) slightly better than a zero vector (error = 1).

\begin{table}[htb]
\caption{Performance evaluation of BEM, random selection method, and our radiation solver for small spatial range radiation. The experiment configuration is consistent with \autoref{table:AcousticNet_result} in the main text.}
\begin{tabular}{lccc}
\hline
\multirow{2}{*}{Method} & \multirow{2}{*}{\begin{tabular}[c]{@{}c@{}}Normalized \\ FFAT Map MSE\end{tabular}} & \multirow{2}{*}{Log Norm MSE} & \multirow{2}{*}{Time} \\
 &  &  &  \\ \hline
BEM & 0 & 0 & 88s \\
Random Selection & 0.68 & 4.68 & 0s \\
Ours & 0.08 & 0.08 & 0.04s \\ \hline
\end{tabular}
\label{table:raidation_small_range}
\end{table}

\subsection{Scalar-valued FFAT Map}
\label{sec:formulation_acoustic}
FFAT Maps can perform fast transfer rendering \cite{sigcourse} and can be efficiently compressed~\cite{kleinpat}.
\begin{figure}[thb]
\centering
\includegraphics[trim={0cm 3.5cm 0cm 3cm},clip,width=0.6\linewidth]{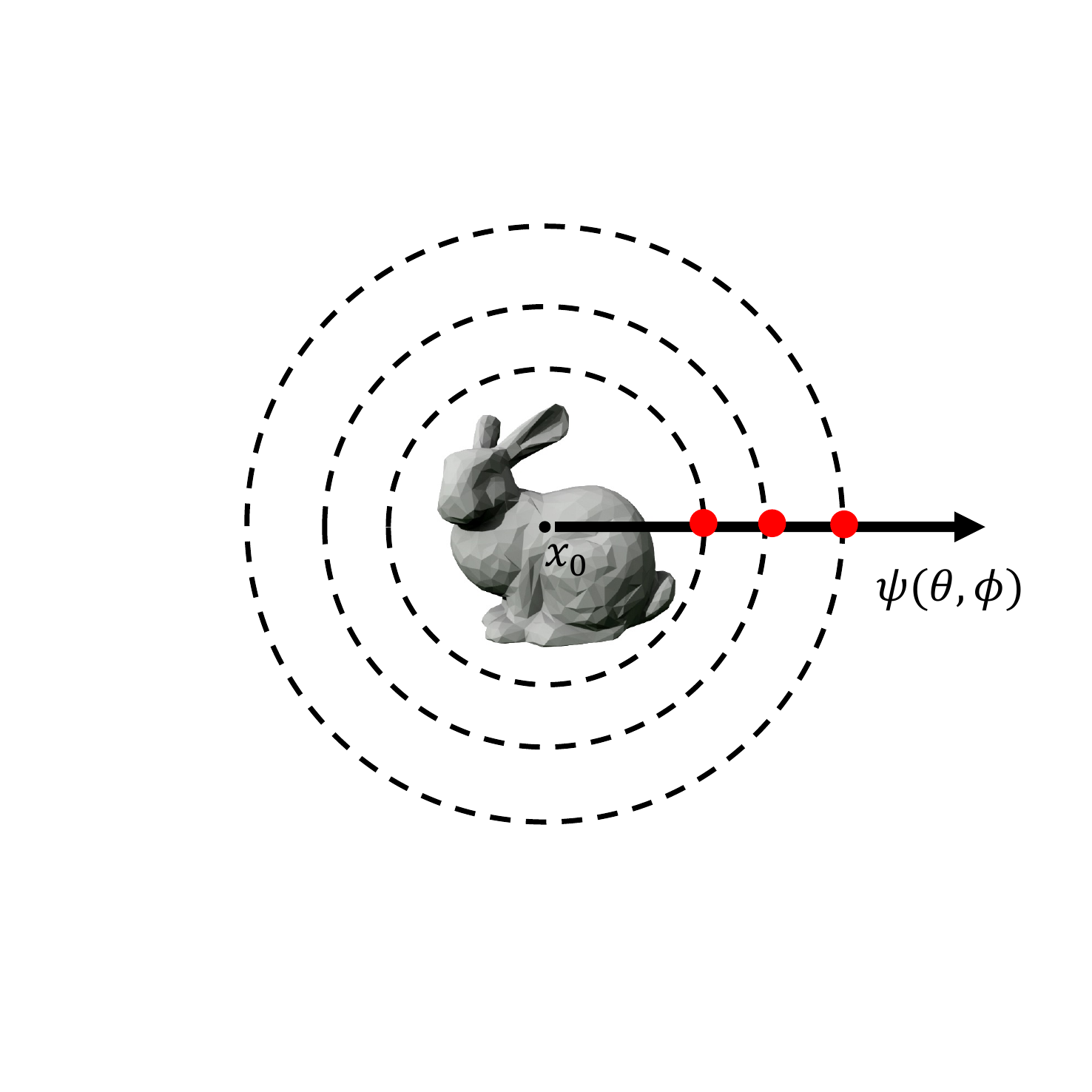}
\caption{Illustration of the scalar-valued FFAT Map. The values in the map are estimated by the least square method for the sample points.}
\label{fig:ffat}
\end{figure}

Our scalar-valued FFAT Map is defined as follows. Acoustic transfer function $p_i(\mathbf{x})$ describes the sound pressure at position $\mathbf{x}$ generated by the vibration of the $i$th mode with unit amplitude. We use a function $\psi_{i}$ to approximate the directionality information of $|p_i(\boldsymbol{x})|$, and we use $1/r$ to approximate the attenuation of the transfer amplitude as the radial distance grows, as shown in \autoref{eq:ffat_map}. $\theta, \phi$ are the coordinates in the spherical coordinate system. When we shoot a ray from $x_0$ at an angle $(\theta, \phi)$, this ray will intersect with $N_s$ spheres with different radii. The value of $\psi_{i}(\theta,\phi)$ can be determined by the least square method based on the values of these intersection points (see \autoref{fig:ffat}). 
We set $N_s = 3$, and the radii of these three spheres are computed as $R_i = (3a)^i, i=1,2,3$, where $a$ is the longest side of the bounding box.
The scalar-valued FFAT Map $\mathbf{\Psi}$ can be computed via a uniform sampling of $(\theta, \phi)$ space and used as the ground-truth for our radiation network solver.

\subsection{Near Spatial Range Radiation}
\label{sec:near_radiation}
As mentioned in Sec. \ref{sec:formulation_acoustic}, our scalar-valued FFAT Maps are computed with three spheres of radii $R_i = (3a)^i, i=1,2,3$, where $a$ is the longest side of the bounding box. Our radiation solver can also be trained for FFAT Maps with different radii. To validate the generalization of our network, we generate the FFAT Map dataset with sphere radii $R_i = (1.25a)^i, i=1,2,3$ and retrain our radiation solver. \autoref{table:raidation_small_range} shows the results of our solver, BEM, and a random selection method (baseline). Our approach also works well for relatively near range radiation and outperforms the baseline significantly in terms of accuracy. We also show comparable results with the ground truth in the accompanying video.

\end{document}